\documentclass[11pt]{article}

\usepackage[T1]{fontenc}
\usepackage[utf8]{inputenc}
\usepackage{booktabs,circledsteps}

\usepackage{epigraph}
\makeatletter
\renewcommand{\epigraph}[2]{%
  \vspace{1ex}%
  \noindent\hfill
  \begin{minipage}{\epigraphwidth}
  \small
    \noindent #1\par
    \vspace{0.5ex}%
    \noindent #2%
  \end{minipage}%
  \vspace{0ex}%
}
\makeatother
\usepackage[margin={0.75in},centering]{geometry}
\usepackage{mathpazo}     
\usepackage{amsmath,bm,amssymb,amsthm}     
\usepackage{microtype}    
\usepackage{enumitem}  

\usepackage{tikz}
\usetikzlibrary{calc, trees, positioning, arrows}

\newtheorem{proposition}{Proposition}

\newtheorem{corollary}{Corollary}
\newtheorem{theorem}{Theorem}

\newtheorem{lemma}{Lemma}
\newtheorem{claim}{Claim}

\theoremstyle{definition}
\newtheorem{definition}{Definition}
\newtheorem{Example}{Example}

\newcommand{\footautoref}[1]{\hyperref[#1]{Footnote~\ref*{#1}}}

\newcommand{\rat}[1][]{%
  \ensuremath{%
    R
  }%
}

\makeatletter
\newcommand{\lk}[1][]{%
  \@ifnextchar^%
    {\lk@sup{#1}}%
    {\ensuremath{%
       L%
       \if\relax\detokenize{#1}\relax
       \else
         _{#1}%
       \fi
       [p]%
     }%
   }%
}

\def\lk@sup#1^#2{%
  \ensuremath{%
    L%
    \if\relax\detokenize{#1}\relax
    \else
      _{#1}%
    \fi
    ^{#2}[p]%
  }%
}
\makeatother

\usepackage{graphicx}
\usepackage{longtable,booktabs,tabularx}
\usepackage[flushleft]{threeparttable}
\usepackage{multirow,array,caption}
\usepackage{pdflscape}      
\usepackage{color,colortbl} 
\usepackage{subcaption}

\definecolor{darkblue}{rgb}{0.15,0,0.37}
\definecolor{darkred}{rgb}{0.35,0,0.08}
\definecolor{mygrey}{rgb}{0.85,0.85,0.85}

\usepackage[authoryear]{natbib}
\usepackage[
pagebackref=false,
breaklinks=true,
colorlinks=true,
linkcolor=black,
citecolor=black,
urlcolor=black,
pdfstartview=FitV,
unicode=true
]{hyperref}



\usepackage[onehalfspacing]{setspace}
\usepackage[multiple]{footmisc}

\usepackage{fancyhdr}
\usepackage{lastpage}

\newcommand{\ShortTitle}[1]{\def\theShortTitle{#1}}
\newcommand{\ShortAuthors}[1]{\def\theShortAuthors{#1}}

\newcommand\blfootnote[1]{%
	\begingroup
	\renewcommand\thefootnote{}\footnote{\hspace*{0pt}#1}%
	\addtocounter{footnote}{-1}%
	\endgroup
}

\newenvironment{niceabstract}
{   \vspace{-2.85em}
	\begin{quote}
    \normalsize
		\setlength{\parskip}{0em}
		\setlength{\parindent}{0em}
	}
	{
	\end{quote}
}

\usepackage{etoolbox}

\makeatletter
\newtoggle{inappendix}
\togglefalse{inappendix}

\pretocmd{\appendix}{\toggletrue{inappendix}}{}{}

\renewcommand{\@seccntformat}[1]{%
  \iftoggle{inappendix}{%
    \ifstrequal{#1}{section}{Appendix \csname the#1\endcsname.\quad}{\csname the#1\endcsname\quad}%
  }{%
    \csname the#1\endcsname\quad
  }%
}
\makeatother

\newcommand{\PaperTitle}[1]{\def\thePaperTitle{#1}}

\newcommand{\AuthorOne}[5]{%
	\def\authorOneFirstName{#1}%
	\def\authorOneLastName{#2}%
	\def\authorOneAff{#3}%
	\def\authorOneLoc{#4}%
	\def\authorOneEmail{#5}%
}

\newcommand{\AutoShortAuthors}{%
    \ShortAuthors{\authorOneFirstName~\authorOneLastName}%
}

\newcommand{\printTitleAndAuthors}{
	\thispagestyle{empty}
	\begin{center}
		{\Large\thePaperTitle}\\[1em]
	\end{center}
	
	\begin{center}
	\begin{minipage}[t]{0.6\textwidth}
		\centering
		{\normalsize \authorOneFirstName\ \authorOneLastName}\\[-0.5em]
		\vspace{4pt}
		{\small
			\authorOneAff \\[-0.33em]
			\authorOneLoc \\[-0.33em]
			\href{mailto:\authorOneEmail}{\texttt{\authorOneEmail}}
		}
	\end{minipage}
	\end{center}
		
	\vspace{0.5em}
	\begin{center}
		{\normalsize \today}
	\end{center}
}

\begin{document}
	
	\ShortTitle{Local Knowledge and the Scale of Markets}
	
	\PaperTitle{How Much Must Individuals Know? \\Local Knowledge and the Scale of Markets}
	
	\AuthorOne{Shuige}{Liu}
	{Center For Mathematical Economics, Bielefeld University}
	{Bielefeld, Germany}
	{shuige.liu@uni-bielefeld.de}

	\AutoShortAuthors
	
	\printTitleAndAuthors
\blfootnote{Early versions of this paper had been circulated under the titles ``Knowledge and Unanimous Acceptance of Core Payoffs: An Epistemic Foundation for Cooperative Game Theory'' and ``Measuring Information Burden: From Coalition-Based Reasoning to the Price System''. The author is grateful to Pierpaolo Battigalli, Giacomo Bonanno, Luca Braghieri, Justus Preusser, Niccol\`{o} Urbinati, and Rakesh Vohra for their detailed comments. She thanks Christian W. Bach, Lorenzo Bastianello, Gerrit Bauch, Chih Chang, Pietro Dall'Ara, Meng-Jhang Fong, Manuel Förster, Satoshi Fukuda, Yukihiko Funaki, Peter Gibbard, Pierfrancesco Guarino, Mamoru Kaneko, Nenad Kos, Fabio Maccheroni, Suraj Malladi, Massimo Marinacci, Zsombor Z. M\'{e}der, Antonio Penta, Andr\'{e}s Perea, Ronald Peeters, Paolo Pin, Rasto Ratislav, Frank Riedel, Marten Ritterath,  Pietro Salmaso, Enrico Mattia Salonia, Burkhard Schipper, Christoph Schottm\"uller, Fernando Vega-Redondo,  Murat \"Ungor, Yanjing Wang, Y\`{\i} N. W\'{a}ng, and participants of 13th LOFT Conference, IV Italian Junior Theory Workshop, seminars in Bocconi University, University of Cologne, and Bielefeld University for valuable discussions and encouragements. The paper was partially supported by Grant-in-Aids for Young Scientists (B) of JSPS No.17K13707 and Grant for Special Research Project No. 2018K-016 of Waseda University. This paper is dedicated to the late Professor Chih Chang of National Tsinghua University, who first inspired the author’s interest in game theory. His untimely passing in 2020 was a profound loss to the field and to all who knew him.}





\begin{niceabstract}
\textbf{\normalsize Abstract.}
How much must individuals know for a large economy to be stable? We study this question in a proof-theoretic framework where information is a set of economically meaningful coalition-feasibility premises, and an agent rejects a proposed allocation only if she can establish from her premises that a profitable coalitional deviation exists. We characterize the distributions of information under which the allocations left unopposed are exactly the core, which defines a minimum informational burden. In Shapley’s (\citeyear{s75}) two-type replica economy, within the \cite{ds63} convergence tradition, the number of coalitions can grow exponentially, yet the minimum burden is only $2k+1$ premises for $k\ge 3$. This is asymptotically one premise per agent. The premises can be distributed so that no agent holds more than two, regardless of economy size. Moreover, every informational covering must contain at least one almost-grand coalition of each type imbalance. Thus, global coalitional stability can be supported by a sparse distribution of local knowledge, but cannot dispense entirely with economy-wide coalition information. Market scale and individual informational scale need not grow together, even though no prices are assumed or derived.
\end{niceabstract}
\begin{center}
		\begin{minipage}{0.9\textwidth}
		\end{minipage}
	\end{center}

\newpage
\begin{spacing}{0.95}
   \tableofcontents
\end{spacing}
\newpage

\epigraph
  {%
    ``But there is a further question which seems to me to be at least equally important, but which appears to have received no attention at all, and that is how much knowledge and what sort of knowledge the different individuals must possess in order that we may be able to speak of equilibrium. It is clear that if the concept is to have any empirical significance it cannot presuppose that everybody knows everything.''%
  }
  {---F. A. Hayek, ``Economics and Knowledge'' (1937)}



\section{Introduction}\label{sec:int}

A modern market connects millions of people, and no participant surveys more than a sliver of it. 
A trader may know a few customers and suppliers, a household only a small set of possible exchanges. Yet their individual decisions form part of a system in which resources can be allocated and traded on a scale far beyond any participant's direct field of knowledge. Since Adam Smith, the capacity of markets to coordinate a multitude of separate activities has been a central concern of economics. \cite{fh37, fh45} placed the dispersion of knowledge at the center of this problem: the relevant circumstances are known to different people, often only locally, while economic coordination is a property of the system as a whole. The fundamental question is therefore not merely how much information a market can process, but how little its individual participants may need to know for a large market to remain viable.

This question becomes especially compelling when an economy expands. Wider markets need not arise as fully formed systems. Historical accounts describe the development of impersonal trade through overlapping local relationships and institutions that made exchange beyond established communities possible \citep[see, for example,][]{ga06}. As the number of potential trading partners increases, so does the number of conceivable groups that might benefit from exchanging or reallocating resources. It does not follow, however, that every participant must learn about a comparable number of new opportunities. Can an economy become more integrated, and its stable allocations more tightly disciplined, without requiring each individual to acquire an ever larger stock of information? What informational relationships would make such expansion possible?

The classical theory of the core provides a precise setting in which to investigate this problem. An allocation belongs to the core if no coalition of agents can improve upon it by reallocating the resources available to its members. The definition is global: in principle, it refers to every possible coalition. In the replica economies of \cite{ds63}, this requirement becomes increasingly demanding as the economy grows, and the core converges toward competitive equilibrium. Replication thus brings together two features of an expanding market: a rapidly increasing collection of possible coalitional deviations and a progressively sharper notion of stability. With $2k$ agents, there are $2^{2k}-1$ nonempty coalitions. Does the informational requirement for recognizing and sustaining the core have to grow at a comparable rate? Or can a small collection of locally distributed facts rule out every allocation outside the core?

This paper establishes the latter possibility. In a two-type replica economy of the kind studied by \cite{s75}, the minimum number of distinct coalition-feasibility facts sufficient for decentralized individual rejection decisions to recover the core is $2k+1$ for every $k\geq 3$. Thus, while the number of possible coalitions grows exponentially, the minimum stock of relevant coalition information grows \emph{linearly} with population. Its average per agent is
\[
\frac{\mu(E_k)}{|N_k|}=\frac{2k+1}{2k}=1+\frac{1}{2k}\longrightarrow 1.
\]
More strikingly, the information can be assigned to coalition members so that \emph{no individual needs to hold more than two coalition-feasibility premises}, independently of market size. An economy's scale and the scale of an individual's informational endowment need not grow together.

To state and prove this result, we make explicit both what agents know and what they can establish from that knowledge. Consider an allocation proposed to a population of agents. An individual has a reason to reject it if a coalition containing her could feasibly reallocate its resources in a way that benefits her without making any other member worse off. The existence of such a deviation is an objective property of the economy. The individual's ability to identify and justify it is a different matter: she may lack the relevant information about what the coalition can achieve. A failure to object need not mean that the proposal is stable; it may simply mean that no one can establish a profitable objection from the information at hand. We therefore ask how coalition-feasibility information must be distributed across agents so that every noncore allocation encounters at least one justified rejection, while no core allocation does.

Our formal framework represents information by sentences with explicit economic meanings and individual reasoning by derivations in Gentzen's (\citeyear{gg35a, gg35b}) sequent calculus. We first develop the framework for transferable-utility games, where elementary sentences assert the feasibility of particular payoff vectors for particular coalitions. An agent rejects a proposed payoff vector only if she can derive the existence of a profitable coalitional deviation from her informational premises. This separates truth from derivability and allows us to identify which premises actually do the inferential work. \autoref{theorem:load-bearing} establishes a necessary and sufficient informational condition for decentralized rejection decisions to recover the core: for every noncore payoff vector, at least one coalition that blocks it must be known to at least one of its members. Equivalently, the coalition information distributed among agents must cover the blocking opportunities of every noncore outcome. Since a single coalition may block many different outcomes, and a noncore outcome may have many potential blockers, an informational covering may be far smaller than the collection of all coalitions.

We call the smallest cardinality of such a covering the \emph{minimum informational burden}, denoted by $\mu$. This is a measure of the stock of distinct coalition-feasibility facts sufficient for a particular collective task, not a count of all logically possible deviations. We extend the characterization to pure-exchange economies, using a first-order language to accommodate their infinite allocation spaces while retaining finite informational premises. In this extension, information that a coalition can feasibly redistribute its endowment serves as one coalition-specific premise, from which further feasibility claims can be derived. The resulting measure permits us to ask not only \emph{how many} coalition facts are needed, but also \emph{which} coalitions they concern and \emph{where} that information can reside.

\autoref{th:info-replica} answers these questions in Shapley's (\citeyear{s75}) two-good, two-type replica economy. Its constructive content is as revealing as its exact bound. For every $k\geq 2$, consider two families of \(k\) cross-type pairs whose union forms one alternating cycle. 
Every agent has exactly two neighbors in this informational network, both of the opposite type. Add two further almost-grand coalitions, each containing all but one of the economy's $2k$ agents, with the omitted agents belonging to different types. These $2k+2$ coalitions form an informational covering, denoted by $\mathcal L_k$. For $k\geq 3$, replacing two disjoint bilateral coalitions by their four-person union yields a second covering, $\mathcal F_k$, consisting of $2k+1$ coalitions. \autoref{th:info-replica} proves that $\mathcal F_k$ attains the minimum. For $k=2$, the cyclic covering $\mathcal L_2$, with six coalitions, is minimum.

Why can so few facts suffice? The construction separates two informational tasks. The balanced coalitions---in the simpler covering $\mathcal L_k$, the bilateral links of the cycle---discipline efficiency and equal treatment. Because the local constraints overlap and the network connects the two types throughout the population, a feasible allocation that none of these coalitions can block must give identical agents identical bundles of a particular form. Once allocations have been reduced to this equal-treatment family, only the division of resources between the two types remains to be determined. The two almost-grand coalitions impose the respective bounds on this division, thereby identifying exactly the replica core. Small, overlapping informational neighborhoods and a fixed number of economy-wide coalition facts perform different, complementary tasks.

Importantly, the two almost-grand coalitions are not merely convenient features of our constructions. \autoref{th:info-replica} further establishes that every informational covering must contain at least one almost-grand coalition excluding an agent of each type. Thus, although the information needed to recover the core can be distributed sparsely across the population, some economy-wide coalition information is indispensable. Local information and global information are not simply alternative means of sustaining stability; they perform complementary functions, and neither can entirely replace the other in this economy.

This necessary complementarity suggests a view of decentralized coordination that is neither purely local nor centrally omniscient. 
In the cyclic construction, ordinary bilateral relationships do not become more numerous for any agent as the economy expands. The only informational objects whose \emph{scope} expands with population are the two almost-grand coalitions. Their feasibility information could, as an economic interpretation, be held or certified by agents performing a coordinating or public role. Such a role is not that of an all-knowing planner: the construction requires a limited set of system-level facts, not a complete description of every feasible transaction. Nor does the theorem require a government, a central institution, or even a unique holder of those facts. What it demonstrates is that global coalitional stability is compatible with an informational division of labor between sparse local relationships and a thin system-level layer.

The result has a further implication for market expansion. A larger replica economy adds potential partners and coalitions, yet its sufficient local network need not be rebuilt from scratch. In the cyclic construction, the entry of a new pair of agents can be accommodated by inserting a few bilateral links and updating the two system-level coalitions, leaving the great majority of existing local relationships unchanged. This is a property of the constructed informational coverings, not a model of actual entry or learning. Nevertheless, it illustrates how an informational architecture supporting global stability can remain sparse and be locally reconfigured as the economy grows. The informational reach of the entire system can increase without a corresponding expansion of each participant's direct informational neighborhood.

Together with the Debreu-Scarf (\citeyear{ds63}) theorem, this yields a different interpretation of the relationship between market size and competitive equilibrium. The classical result shows that replication brings the core toward competitive equilibrium. Our result shows that this movement toward competitive stability need not require a growing individual stock of coalition-feasibility information. Competitive market order can therefore be approached, in this precise informational sense, through globally sufficient but individually limited knowledge. The result does not establish that bilateral trading relations generate prices, that an informational network arises endogenously, or that the resulting allocation is reached through a decentralized adjustment process. These are questions of dynamics and institutional formation. Instead, the theorem identifies an informational condition under which their possibility is not defeated by the increasing number of coalitional alternatives.

There are two important qualifications to the interpretation. First, informational sufficiency is assessed relative to a specified decision problem: agents independently justify objections to proposed allocations on the basis of coalition feasibility. We do not minimize communication over arbitrary protocols, nor do we establish a price-discovery mechanism. Second, the measure counts economically interpretable coalition-specific premises, not their description length in bits, the cost of acquiring them, or the length of the proofs that use them. An almost-grand coalition is one informational unit in our measure even though the scope of the corresponding fact grows with the economy. The substantive finding is therefore not that all information-processing or organizational costs remain bounded, but that the \emph{number and distribution of logically sufficient coalition-feasibility premises} can remain exceptionally sparse. The distinction matters both for what the theorem establishes and for the institutional questions it opens.

The analysis also contributes a methodological approach to the economics of information. As \cite{lh69} emphasized, a count of symbols has little economic content unless their meanings are specified. Our informational units are statements about feasible economic actions, and our proofs make explicit which of those statements are necessary for a particular decision. This makes it possible to distinguish the amount of information potentially available in an economy from the amount actually needed for a collective economic property to be recognized. Beyond the core, the same distinction suggests a way of evaluating institutions and mechanisms by the informational and inferential demands they place on participants. The question is not only what an institution achieves, but what it requires individuals to know in order for their decisions to sustain that achievement.

The rest of the paper is organized as follows. Section \ref{sec:p}  reviews transferable-utility games and the core. Section \ref{sec:logic} introduces the formal language and proof-theoretic representation of information and individual reasoning. Section \ref{sec:cor} establishes the informational characterization of the core and defines minimum informational burden. Section \ref{sec: DS} extends the framework to pure-exchange economies and, using the Debreu-Scarf replica setting, establishes the sparse informational foundations of the replica core. The proofs are collected in the Appendix.


\subsection{Contribution and Related Literature}

\paragraph{Local knowledge, prices, and market order.}
Our point of departure is Hayek's (\citeyear{fh37, fh45}) understanding of market coordination as an informational problem. In his account, individuals act on fragments of knowledge, while local adjustments may be coordinated across a much larger economic system. This is not merely a claim that an equilibrium price vector is a compact statistic: it is also a claim about the relation between limited individual fields of knowledge and economy-wide order. \cite{bks17} emphasize this local-to-global dimension of Hayek's argument, while distinguishing the insight that markets process dispersed information from the stronger claim that decentralized markets must therefore perform well in every circumstance. Our paper isolates a mathematical question suggested by the former insight. Rather than presupposing prices as coordinating signals, we ask how sparse the distribution of directly interpretable coalition-feasibility information can be while still sufficing to distinguish globally stable from unstable allocations.

The distinction matters for understanding market formation as well as market operation. Menger's (\citeyear{mk1892}) account of the origin of money and Young's (\citeyear{yp93}) analysis of the evolution of conventions study, in different ways, how institutional regularities can emerge from decentralized behavior. Economic history has also stressed that wider exchange depends on social and institutional arrangements rather than on an abstract collection of isolated bilateral trades. \cite{ga06}, for example, examines institutional support for impersonal exchange among merchants who could not rely on direct knowledge of their trading partners. These works concern the emergence, enforcement, or evolution of institutions, whereas we take the informational structure as given and derive a static sufficiency and minimality result. Their relevance is to the question the theorem brings into focus: how can the informational reach of exchange expand beyond the direct knowledge of its individual participants? Our answer is deliberately partial, but it shows that the global informational task need not impose a correspondingly global burden on each person.

A substantial literature following Hayek investigates what information prices communicate or aggregate, beginning with \cite{rr68, rr72, rr79} and including \cite{gs80} and \cite{hw80}, with more recent contributions such as \cite{cyz22, aht23}, and \cite{gik24}. A related literature investigates the strategic foundations of competitive or efficient market outcomes under private information \citep{w77, m79, m81, bgw24}. Those approaches typically specify informational or strategic environments and analyze the functioning of prices or market mechanisms. We instead fix a collective stability criterion and ask which informational premises must be available for decentralized individual objections to recover it. Thus prices motivate our inquiry, but their informational content, existence, and adjustment are not the objects measured by our theorem.

\paragraph{Core convergence and the informational structure of stability.}
\cite{ds63} establish the connection between expanding replica economies and competitive equilibrium. \cite{s75} provides a particularly tractable characterization of the core in the two-type exchange economy we study. These classical results describe which allocations survive coalitional objections as replication proceeds. Our question is different: what is the smallest collection of coalitions whose feasibility information suffices to expose every allocation that does not survive? The difference between a characterization of the core by all relevant inequalities and a \emph{minimum informational covering} is essential. Many coalitional restrictions may be logically redundant once other restrictions are present. Theorem~2 identifies that redundancy exactly in Shapley's economy and, beyond obtaining a linear minimum, constructs a distribution of the requisite premises with uniformly bounded individual burden. It thus supplies an informational complement to the classical convergence result. \autoref{th:info-replica} also identifies an irreducible economy-wide component: every informational covering must contain two almost-grand coalitions of opposite type imbalances, even though the overall informational stock can remain sparse.

\paragraph{Reducing the family of coalitions needed to determine a coalitional solution.}
Our analysis is related to a long-standing question in cooperative game theory: how much of the coalitional structure of a game is needed to characterize its solutions? One strand of the literature identifies coalitions that play a particularly important role in determining cooperative solutions, beginning with the vital coalitions of \cite{gd59} and subsequently developing notions of essentiality \citep{hg80, bst05, ss16, dp24} and relevance \citep{bf16}. Another strand investigates the structural properties of collections of coalitions and their implications for core existence and characterization \citep{kw82, bgv97, dr98}. Closely related is the computational literature on characterization sets for the nucleolus \citep{ggz98, rp98}, which exploits the redundancy of coalitional restrictions to reduce the complexity of computing cooperative solutions.

While this paper share with this literature an interest in coalitional redundancy, our motivation is cognitive rather than structural or computational. We ask not which coalitions possess particular structural properties, nor how efficiently a cooperative solution can be computed, but what individuals collectively need to know for their separate judgments to sustain that solution. From this perspective, information about a structurally redundant coalition may nevertheless be valuable, depending on what other information is available in the population. This leads us to formulate informational sufficiency as a covering problem, in which coalitional information is evaluated through its joint capacity to support individual rejection decisions. Theorem 1 provides the foundation for this formulation, while Theorem 2 identifies the minimum informational burden in a replica economy, its distribution across individuals, and the kinds of coalitional information that are indispensable.

\paragraph{Restricted blocking in exchange economies.}
A related literature investigates the extent to which blocking can be restricted without losing the economic implications of the core. Restrictions considered include the size and composition of blocking coalitions \citep{sd72, gb72, vk72, ma79, be23}, as well as the communication structures through which coalitions can form \citep{gw83, os85, kow86}. A complementary strand investigates the abundance of blocking coalitions in large economies \citep{ma78, gwy07}.\footnote{The literature on restricted cooperation is also connected to the broader literature on networks in cooperative games \citep{mrb77} and endogenous network formation \citep{jw96, bg00, km01}. Unlike these approaches, our analysis holds the underlying possibilities of cooperation fixed and investigates restrictions on individuals' information about these possibilities.} A further strand of the literature studies whether the gains from cooperation can be realized, approximately or exactly, by coalitions bounded in size \citep{wm92, ww08}.

While sharing the informational concerns underlying this literature, we ask not what kinds of coalitions suffice to characterize the core, but how few coalitions suffice. Our measure of informational burden concerns the number of distinct pieces of coalitional information required, rather than the cost of acquiring or processing each piece. Accordingly, we impose no \emph{a priori} restriction on the size or composition of the coalitions constituting an informational covering. The structure of a minimum covering is thus an outcome of the analysis. In Shapley's replica economy, a minimum covering can be constructed from small, locally organized coalitions and two indispensable almost-grand coalitions.

\paragraph{Informational requirements and communication complexity.}
\cite{lh73} proposed that economic allocation procedures be evaluated in terms of their informational and computational requirements as well as their outcomes. \cite{mr74} pursue one formulation through the size of message spaces in decentralized allocation mechanisms; \cite{sr74} relates such measures to information revealed by allocation processes. The subsequent communication-complexity approach measures the messages that parties must exchange to compute or verify a desired outcome \citep{kn97-2}. \cite{ns06} show, in particular, that communication requirements for efficient allocation and its verification can be large, and relate efficient allocation to supporting prices. We measure a different resource: the minimum stock of coalition-feasibility premises that can be distributed among agents so that their own justified objections collectively characterize the core. Our minimization is over such informational coverings, not over communication protocols. Consequently, the linear bound in Theorem~2 does not contradict communication lower bounds, and the bounded number of premises per agent should not be read as a bound on their bit complexity.

This distinction also bears on the informational evaluation of mechanisms. \cite{ls17} studies the simplicity of mechanisms from the perspective of agents' contingent reasoning. Our framework likewise treats the demands imposed on actual decision-makers as part of the economic object to be evaluated, but it proceeds by fixing the meanings of elementary information and determining what can be derived from them. In particular, we distinguish a mechanism's performance from the informational stock required for agents to justify the decisions that support a specified outcome. The replica result provides a concrete example in which the organization and distribution of information matter as much as its aggregate amount.

\paragraph{Incomplete information and cooperative stability.}
The literature on the core under incomplete information, including \cite{w78, v99, dv05, sv07, dg07}, and \cite{m07}, investigates how limited information and beliefs affect cooperation and coalitional stability. Our approach takes the complete-information core as a target and asks which information is sufficient for individual decisions to recover it. Serrano and Vohra's (\citeyear{sv07}) emphasis on individual decisions is especially close in spirit. The difference is that we represent an informational gap by the absence of a derivable justification: a coalition may in fact block an allocation without any of its members being able to establish this from her premises. In this way, ignorance is not automatically replaced by a probabilistic belief, and the question of information distribution becomes an explicit covering problem.

\paragraph{Formal language and proof theory.}
The syntactic treatment of information connects the paper to formal epistemic game theory, including \cite{b08, b15} (for surveys, see \cite{fhmv95, bb99}). Proof-theoretic approaches to games have been developed by \cite{kn96, kn97} and subsequent work. Our use of proof theory is directed toward measurement: explicit derivations reveal which economically meaningful premises are load-bearing for an individual's rejection decision, while the underlying logic distinguishes what is true of the game from what is provable on the basis of a particular informational endowment. The proof-theoretic apparatus is therefore not an alternative way of stating the core; it enables the definition and computation of the informational burden that culminates in \autoref{th:info-replica}.

\section{TU Games and the Core: An Overview}\label{sec:p}

This subsection provides an overview of cooperative games with transferable utility, as well as the concept of the core. For a more detailed treatment, see \cite{k92} and Peleg and Sudhoelter (\citeyear{ps07}, Sections 2--3).


A \emph{cooperative game with transferable utility (TU game)} is a pair $G=(N,v)$, where $N$ is the finite \emph{set of agents (players)} and $v:2^{N}\rightarrow \mathbb{R}_+$ with $v(\emptyset )=0$ is the \emph{characteristic function}.\footnote{In this paper, we let $\mathbb{R}_+:=[0, \infty)$, $\mathbb{N}: =\{1,2,...\}$, and $\mathbb{N}_0:=\mathbb{N} \cup \{0\}$. For simplicity, we restrict attention to nonnegative characteristic functions; allowing negative values does not affect the analysis.}  For each \emph{coalition} $S \subseteq N$, $v(S)$, called the \emph{value} of $S$, is interpreted as the maximal total payoff that can be obtained by $S$ (for example, through their  collective action). Given a TU game $G = (N, v)$, a \emph{payoff vector} is a real-valued vector $x = (x_i)_{i \in N} \in \mathbb{R}_+^N$, where each $x_i $ represents the amount of income received by player $i$. For each $S \subseteq N$, $x$ is \emph{feasible} for $S$ if $\sum_{i\in S}x_i \leq v(S)$.

A central question in cooperative game theory is which payoff vectors, or distribution plans, should be regarded as acceptable given the structure of the game. A natural requirement is that an acceptable payoff vector be sustainable within the grand coalition $N$. This leads to two conditions. First, the payoff vector should be efficient, in the sense that it allocates the value generated by $N$. 
Second, no subset of players should have an incentive to deviate: there should be no coalition that can feasibly achieve an alternative outcome making all its members at least as well off and at least one strictly better off.
These two conditions together define the \emph{core}, a canonical solution concept in cooperative game theory.

Formally, given $x,y\in \mathbb{R}^{N}$ and $S\subseteq N$, we say that $y$ \emph{dominates} $x$ \emph{via} $S$, denoted by $y$ dom$_{S}$ $x,$ iff the following conditions hold: (i) $y_{i}\geq x_{i}$ for each $i\in S$, and (ii) $y_{i}>x_{i}$ for some $i\in S$.\footnote{When only (i) holds, we say that $y$ \emph{weakly dominates} $x$.}




\begin{definition}[\textbf{The core}]\label{def1}
Given $G=(N,v).$ The \emph{core} of $G$, denoted by $C(G)$, is the set of payoff vectors $x$ satisfying the following two conditions:
\begin{itemize}
\item[] \textbf{Efficiency}: $\Sigma _{i\in N}x_{i}=v(N)$;

\item[] \textbf{Coalitional Stability}: There is no $S\subseteq N$ and $y\in \mathbb{R}^{N}$ such that $\Sigma_{i\in S}y_{i}\leq v(S)$ and $y$ dom$_{S} x$.
\end{itemize}
Each $x \in C(G)$ is called a \emph{core payoff vector}.
\end{definition}


%




It can be shown that a payoff vector is coalitionally stable if and only if it satisfies
\begin{equation}
\Sigma _{i\in S}x_{i}\geq v(S) \text{ for each } S\subseteq N
\tag{\(\star\)}
\end{equation}
Therefore, the core can be alternatively defined by Efficiency and Condition (\(\star\)).\medskip
Because our framework is finitistic, we restrict attention to TU games with integer-valued coalition payoffs and to payoff vectors with integer coordinates. Formally, we consider games \(G=(N,v)\) such that \(v(S)\in \mathbb N_0\) for every \(S\subseteq N\), and for each such game we study payoff vectors in \(\textsf{PV}(G):=\{x \in \mathbb{N}_0^N: x_i \in [0, M_G] \text{ for each } i \in N\}\), where $M_G$ is an integer satisfying \(v(S) \leq M_G\) for each \(S \subseteq N\).\footnote{The number \(M_G\) need not equal \(\max_{S \in 2^N} v(S)\). What matters is that it be a commonly known and accepted upper bound on feasible payoffs. One may think, for example, of \(M_G\) as a publicly recognized aggregate bound, such as annual GDP. We impose such a bound in order to ensure that the criterion \textsf{R}$_i(x)$ introduced in Section~\ref{subsec:criterion} is a finite formula.}
Under this restriction, the relevant solution concept coincides with the usual core restricted to \(\textsf{PV}(G)\): if an integer-valued payoff vector is blocked by some feasible deviation in \(\mathbb R^N\), then it is also blocked by an integer-valued feasible deviation. Thus no generality is lost, at the level of integer-valued payoff vectors, by restricting attention to deviations in \(\textsf{PV}(G)\).\footnote{Indeed, suppose \(x\in \textsf{PV}(G)\) is not in the usual core. Then there exist \(S\subseteq N\) and \(y\in \mathbb R^N\) such that \(\sum_{i\in S} y_i\le v(S)\) and \(y\) dominates \(x\) via \(S\). It follows that \(\sum_{i\in S}x_i < v(S)\). Since both \(\sum_{i\in S}x_i\) and \(v(S)\) are integers, we have \(\sum_{i\in S}x_i\le v(S)-1\). Hence, for any \(j\in S\), the vector \(z\in \textsf{PV}(G)\) defined by \(z_j=x_j+1\) and \(z_i=x_i\) for \(i\neq j\) is feasible for \(S\) and dominates \(x\) via \(S\).} For convenience, we therefore continue to write \(C(G)\) for the set of payoff vectors in \(\textsf{PV}(G)\) that are efficient and coalitionally stable. 

\section{Describing Decision-Making in a Formal Language}\label{sec:logic}

To study information sufficiency, we begin by recasting a TU game as a profile of individual decision problems: given a proposed payoff vector, each agent decides whether to reject it on the basis of the information she possesses. 
We represent information as a finite set of \emph{sentences} in a formal language and reasoning as proof. An agent rejects a payoff vector if and only if there exists a proof establishing that her decision criterion is satisfied by that vector, given the sentences she holds. Our interest is in the informational conditions under which the set of (unanimously) unopposed payoff vectors coincides with the core.

This section builds the framework in two stages. Section~\ref{subsec:example} presents a self-contained motivating example---a two-player TU game---that illustrates all the key concepts concretely before the formal definitions are given. Sections~\ref{subsec:formal1}--\ref{subsec:reasoning} then develop the language and proof system underlying the example. Readers who work through this example will find the formal definitions straightforward, as they simply generalize what the example already shows. 


\subsection{A Motivating Example}\label{subsec:example}


Consider a two-player TU game $G = (N, v)$ with $N = \{1, 2\}$, $v(\{1\}) = v(\{2\}) = 10$, and $v(\{1,2\}) = 30$. The core of $G$ is $\{(x, 30 - x) : 10 \leq x \leq 20\}$. We take player~$1$'s viewpoint, and suppose she is asked whether to reject the proposal $\bar x = (14, 15)$, which lies outside the core.

\medskip
\noindent\textit{Step 1: How to describe relevant information?}

The relevant information about $G$ can be captured as atomic sentences of two kinds. First, \emph{feasibility sentences}: for each payoff vector $y$ and coalition $S$, the symbolic expression (i.e., a sentence) $y^S$ indicates that coalition $S$ can achieve payoff $y_i$ for each $i \in S$ through collective action. For example, $(11, 19)^{\{1,2\}}$ says ``the grand coalition can achieve a payoff $11$ for agent $1$ and $19$ for agent $2$'', while $(9, \cdot)^{\{1\}}$ says ``agent $1$ can achieve payoff~$9$ by working alone''. Second, \emph{dominance sentences}: $y \geq_S z$ indicates the statement that $y$ weakly dominates $z$ via $S$, that is, $y_i \geq z_i$ for all $i \in S$. These are the only two types of atomic sentence in our language. All more complex statements---including the decision criterion itself---are built from them using the standard logical connectives $\lnot$, $\wedge$, $\vee$, $\rightarrow$.



It is important to note that an atomic sentence such as $(14, 15)^{\{1\}}$ is a syntactic object: it represents the
\emph{proposition} that agent~$1$ can achieve payoff~$14$ alone, but it does not assert that this proposition is true. In the game at hand, $(14, 15)^{\{1\}}$ is a \emph{legitimate sentence} but is in fact \emph{false} (because $v(\{1\}) = 10 < 14$) and $(15, 15)^{\{1,2\}}$ is true (because $15+15 = 30 = v(\{1,2\})$). Whether the sentence is true or false in the game is a semantic matter; which sentences an agent can \emph{hold}---that is, be aware of and therefore use in proofs---is a separate question.

\medskip
\noindent\textit{Step 2: What is the decision criterion?}

Although many decision criteria could in principle be considered, our interest here is in one induced by the core. The idea is to translate the core's coalitional stability requirement into individual decision problems: agent \(1\) rejects a proposed payoff vector \(x\) if she can establish the existence of a feasible coalitional deviation that makes her strictly better off without making any other member of the coalition worse off. Since there are only two coalitions containing agent 1, such a deviation must satisfy at least one of the following two sentences:

\begin{equation}\label{exp02:1}
  \bigvee_{y \in \textsf{PV}(G)} \bigl(y^{\{1\}} \wedge (y >_{\{1\}} \bar x)\bigr)
\end{equation}
\begin{equation}\label{exp02:12}
  \bigvee_{y \in \textsf{PV}(G)} \Bigl(y^{\{1,2\}} \wedge
    (y \geq_{\{1,2\}} \bar x) \wedge (y >_{\{1\}} \bar x)\Bigr)
\end{equation}


Sentence~(\ref{exp02:1}) states that agent \(1\) can obtain strictly more than \(x_1=14\) by working alone.
Sentence (\ref{exp02:12}) that the grand coalition can achieve a payoff vector that makes neither agent worse off and agent \(1\) strictly better off. These are the two possible grounds for rejecting \(x\) from agent \(1\)'s perspective. We therefore define her rejection criterion as their disjunction:
\begin{equation}\label{exp02:target}
  \textsf{R}_1(\bar x) \;:=
    \bigvee_{y \in \textsf{PV}(G)} \bigl(y^{\{1\}} \wedge \bigl(y >_{\{1\}} \bar x)\bigr)
    \;\vee\;
    \bigvee_{y \in \textsf{PV}(G)} \Bigl(y^{\{1,2\}} \wedge
      (y \geq_{\{1,2\}} \bar x) \wedge (y >_{\{1\}} \bar x)\Bigr)
\end{equation}

\medskip
\noindent\textit{Step 3: Why axioms alone are not enough?}

To say that agent \(1\) rejects \(x\) is, in our framework, to say that she can derive \(\textsf{R}_1(x)\) from the information available to her.\footnote{Strictly speaking, what is at issue here is not the provability of the formula \(\textsf{R}_1(x)\) by itself, but the provability of a \emph{sequent} (which will be defined below) asserting that \(\textsf{R}_1(x)\) follows from the agent’s information \(\Gamma_1\). See the formal definition in Section \ref{subsec:criterion}. This distinction is important for our purposes: a sequent makes explicit from which information a statement is established, and thus allows us to track the informational basis of a decision. That is one reason for working in a Gentzen-style system rather than a Hilbert-style one.} To establish such a claim of derivability, we need a proof. A proof is a finite tree: it begins from terminal nodes, called axioms, and proceeds upward by legitimate inference rules until it reaches the conclusion to be established. In this paper, we use a version of Gentzen's sequent calculus, whose logical axioms are tautological identities, together with a small collection of inference rules.

Can agent $1$ prove the claim from logical axioms (i.e., tautologies) alone? The answer is no. We need also to include some \emph{non-logical axioms}, which should not themselves encode any such game-specific information but capture only general reasoning capacities that it is natural to attribute to any agent independently of the particular game---for example, the ability to compare payoff vectors coordinate-wise. In fact, the only non-logical axioms in our framework are numerical comparison facts---for instance, that \((8,7)\geq_{\{2\}}(9,2)\) is an axiom because $7 \geq 2$, and \((8,7)\geq_{\{1,2\}}(9,2)\) is not because $8<9$. These are informative enough to tell the agent whether one payoff vector is at least as good as another for a given coalition, but they say nothing about which payoff vectors are feasible for which coalitions.

\medskip
\noindent\textit{Step 4: What information suffices, and how is it used in a proof?}

Suppose agent~$1$ holds the following set of sentences $\Gamma(\mathcal{N}_1)$, which encodes the complete and exact information about true values of $v(\{1\})$ and $v(\{1,2\})$:
\begin{equation}\label{exp02:full_information}
\begin{split}
 \Gamma(\mathcal{N}_1) &:=
    \{y^{\{1\}}   : y_1 \leq v(\{1\})\}
    \;\cup\; \{\lnot y^{\{1\}}   : y_1 > v(\{1\})\} \\
  &\phantom{:=}\;\cup\;
    \{y^{\{1,2\}} : y_1+y_2 \leq v(\{1,2\})\}
    \;\cup\; \{\lnot y^{\{1,2\}} : y_1+y_2 > v(\{1,2\})\}
\end{split}
\end{equation}

We claim that, given this information, agent \(1\) can derive $\textsf{R}_1(x)$; in other words, the information in $\Gamma(\mathcal{N}_1)$ suffices to establish that $x = (14, 15)$ is rejected by agent \(1\). The proof proceeds along the following lines: for some \(y^S \in \Gamma((\mathcal{N}_1)$, $y$dom$_Sx$ and $y_1 > x_1$. For example, consider \(\bar y = (15, 15)\):





\begin{equation}
\frac{
\dfrac{\dfrac{ \bar y^{\{1, 2\}} \Rightarrow \bar y^{\{1, 2\}}, \ \ \ \ \Rightarrow \bar y \geq_{\{1,2\}}\bar x, \ \ \ \ \ \Rightarrow \bar y >_{\{1\}}\bar x}{\bar y^{\{1, 2\}}\Rightarrow \bar y ^{\{1, 2\}} \wedge \Bigl(\bar y \geq_{\{1,2\}}\bar x\Bigr) \wedge \Bigl(\bar y >_{\{1\}} \bar x\Bigr)} \;(\text{Weakening, }\wedge\text{-Right})}
 {\Gamma((\mathcal{N}_1) \Rightarrow \bar y ^{\{1, 2\}} \wedge \Bigl(\bar y \geq_{\{1,2\}}\bar x\Bigr) \wedge \Bigl(\bar y >_{\{1\}} \bar x\Bigr)}(\text{Weakening})}
{\Gamma(\mathcal{N}_1) \Rightarrow \mathsf{R}_1(\bar x)} \;(\vee\text{-Right})
\end{equation}

\smallskip
The proof is read from top to bottom. The three uppermost sequents are the ``leaves'' of the proof tree: they are axioms and require no further justification within the proof system. The first, \(\bar y^{\{1,2\}}\Rightarrow\bar y^{\{1,2\}}\), is an instance of the logical axiom (tautology) stating that a sentence follows from itself. The other two are non-logical axioms, expressing that \(\bar y\) weakly improves both agents' payoffs and strictly improves agent 1's payoff relative to \(\bar x\). Applying the inference rules \(\wedge\)-Right (and Weakening when necessary) we combine these statements.\footnote{Strictly speaking, each step in a formal derivation applies only one inference rule. Here, we abbreviate successive applications of Weakening and \(\wedge\)-Right to save space.} Since \(\bar y^{\{1,2\}}\) belongs to \(\Gamma(\mathcal N_1)\), Weakening allows us to replace the antecedent by the agent's full information set. Finally, \(\vee\)-Right establishes the rejection criterion \(\mathsf R_1(\bar x)\), of which the particular deviation is one disjunct. We thus obtain \(\vdash\Gamma(\mathcal N_1)\Rightarrow\mathsf R_1(\bar x)\): agent 1 can justify rejecting the proposed payoff vector.

\medskip
\noindent\textit{Step 5: The role of ignorance and information distribution.}

So far, agent \(1\) has been endowed with ``full'' information about all coalitions containing her. But what happens when her information is insufficient? Suppose, for instance, that she possesses information about coalition \(\{1\}\) but not about the grand coalition \(\{1,2\}\). The proof above then no longer goes through: without the relevant feasibility information, she is unable to establish the existence of a profitable deviation and hence cannot justify rejecting \(x=(14,15)\). Importantly, the absence of a proof of rejection does not imply that she can establish the negation of the rejection criterion. Her information may simply be insufficient to derive either conclusion. In such a case, she does not reject the proposal, although she can neither establish that no profitable deviation exists.

This observation leads to the collective informational problem that motivates our analysis. Agents may possess different information about feasible coalitional alternatives, so a proposed payoff vector that cannot be rejected by one agent may nevertheless be rejected by another. We call a payoff vector \emph{unopposed} if no agent can justify rejecting it. The question is therefore not whether each agent possesses enough information to reach a definite conclusion, but how coalition-feasibility information must be distributed across agents for the set of unopposed payoff vectors to coincide exactly with the core. This is the central question addressed in Section \ref{sec:cor}.




\subsection{Formal Language: Alphabet and Sentences}\label{subsec:formal1}

We now give the general definitions that underlie the example. 
A \emph{formal language} consists of four components:\footnote{For a thorough introduction to formal languages, inference rules, and proofs, see textbooks such as \cite{eft21} and \cite{mgz21}.} an \textbf{alphabet} of atomic symbols, a set of \textbf{sentences} (well-formed strings of symbols), \textbf{axioms} (claims taken as true without proof), and \textbf{inference rules} (specifying how new claims may be derived from existing ones). These determine what an agent can express, what she takes for granted, and what she can prove.

\paragraph{Alphabet.} The alphabet consists of:
\begin{itemize}
  \item \textbf{Atomic formulae:} $\mathbf{p}_0, \mathbf{p}_1,
    \mathbf{p}_2, \ldots$; the set of atomic formulae is denoted
   by $\mathsf{At}$.
  \item \textbf{Logical connectives:} $\lnot$ (not), $\wedge$ (and),
    $\vee$ (or), $\rightarrow$ (implies).\footnote{Many works, including Gentzen's seminal papers (\citeyear{gg35a,gg35b}), use $\supset$ for implication. We follow \cite{eft21} in using
    $\rightarrow$, which we find more intuitive for economists. Also, note that we include the connective $\rightarrow$ for generality and to keep the presentation of the sequent calculus standard, although the substantive analysis in this paper uses only $\lnot$, $\wedge$, and $\vee$.}
  \item \textbf{Auxiliary symbols:} parentheses $(\ , )$.
\end{itemize}

\paragraph{Sentences.} A \emph{formula} (or a \emph{proposition}) is any expression obtained
by finitely many applications of:
\begin{itemize}
  \item[\textbf{F0.}] Each atomic formula is a formula;
  \item[\textbf{F1.}] If $A$ is a formula, so is $\lnot A$;
      
  \item[\textbf{F2.}] If $A$ and $B$ are formulae, so are $A \rightarrow B$, $A \wedge B$, and $A \vee B$.
\end{itemize}
We use $\mathsf{F}$ to denote the set of all formulae. The atomic formulae are the indivisible building blocks: they represent basic propositions that cannot be decomposed further. Complex propositions are built from them by applying negation and the connectives.

\subsection{The Language for TU Games}\label{subsec:app1}

We now specialize the abstract language of Section \ref{subsec:formal1} to the TU-game setting. As the motivating example has already illustrated, our aim is to represent the information relevant to an agent’s rejection decision regarding a proposed payoff vector. The atomic formulae are therefore chosen to encode the two basic ingredients of that reasoning: coalition feasibility and comparison of payoff vectors.


Given a TU game $G = (N, v)$, we define the set of atomic formulae $\mathsf{At}(G)=\mathsf{At}_\mathrm{fv}(G) \cup \mathsf{At}_\mathrm{wd}(G)$, where
\begin{align*}
  \mathsf{At}_\mathrm{fv}(G) &:= \{x^S : x \in \mathsf{PV}(G),\;
    S \subseteq N\}, \\
  \mathsf{At}_\mathrm{wd}(G) &:= \{x \geq_S y : x, y \in
    \mathsf{PV}(G),\; S \subseteq N\}.
\end{align*}

As mentioned in the example in Section \ref{subsec:example},
$x^S$ represents the proposition that coalition $S$ can achieve payoff $x_i$ for each $i \in S$ through collective action, and $x \geq_S y$ represents the proposition that $x$ weakly dominates $y$ with respect to $S$ (i.e., $x_i \geq y_i$ for all $i \in S$). Both are atomic symbols: they have no internal logical structure and cannot be further decomposed. Their truth value in the game depends on $v$, but an agent may hold or lack them regardless of whether they are true. Sentences in $\mathsf{F}(G)$ are constructed from $\mathsf{At}(G)$ by finitely many applications of rules \textbf{F0}--\textbf{F2}. In particular, we abbreviate $x \geq_S y \wedge \lnot(y \geq_S x)$ as $x >_S y$.

\subsection{Axioms, Inference Rules, and Proof}\label{subsec:form2}

A language specifies what can be \emph{said}; a proof system specifies what can be \emph{derived}. We now introduce the derivation machinery: axioms (sentences accepted without proof) and inference rules (valid derivation steps). Among the various proof systems in the literature--- Hilbert-style systems, natural deduction---we follow \cite{kn97} and adopt \emph{Gentzen's sequent calculus} \citep{gg35a,gg35b}, which has proven particularly powerful for understanding the structure of logical reasoning.\footnote{For general discussions of proof theory and sequent calculi, see, among others, \cite*{nvp08, ts00, mgz21}. For more detailed treatments of sequent calculi in particular, see \cite{ai21, vp12}.}


\paragraph{Sequents.} A \emph{sequent} is an expression $\Gamma \Rightarrow \Theta$, where $\Gamma$ (the \emph{antecedent}) and $\Theta$ (the \emph{succedent}) are finite (possibly empty) sets
of formulae.\footnote{Many works, including Gentzen's (\citeyear{gg35a,gg35b}), use $\rightarrow$ instead of $\Rightarrow$. We follow \cite{mgz21} and \cite{ai21} in using $\Rightarrow$ to avoid confusion with the implication connective.} The intuition is: if all formulae in $\Gamma$ are true, then at least one formula in $\Theta$ is true;  in other words, given $\Gamma = \{A_1, ..., A_m\}$ and $\Theta = \{B_1,...,B_n\}$, $\Gamma \Rightarrow \Theta$ holds if and only if $A_1 \wedge .... \wedge A_m \rightarrow B_1 \vee ... \vee B_n$ holds.\footnote{Although a sequent can be read in terms of an implication formula, it is not merely a notational variant of $\rightarrow$. The connective $\rightarrow$ belongs to the formal language, whereas $\Rightarrow$ belongs to the proof system (i.e., a \emph{meta-language} that makes the role of premises and conclusions explicit). This is precisely why sequents are useful: they display \emph{derivational structure} directly, rather than forcing it to be encoded inside formulae.} When $\Gamma$ is empty, $\ \ \Rightarrow \Theta$ means $\Theta$ holds unconditionally. When $\Theta$ is empty, $\Gamma \Rightarrow \ \ $ means the conjunction of $\Gamma$ is contradictory.

\paragraph{Proofs.} A \emph{proof} is a finite tree of sequents in which each leaf is an axiom and each internal node is obtained from its children by one inference rule. The sequent at the root is the sentence proved. We write $\vdash \Gamma \Rightarrow \Theta$ to indicate that $\Gamma \Rightarrow \Theta$ is provable.

\paragraph{Inference rules.} Following the LK system modified by \cite{k02}, the rules are:

\smallskip
\noindent\textbf{Structural rules:}
\begin{equation*}
\frac{\Gamma \Rightarrow \Theta}
     {\Delta, \Gamma \Rightarrow \Theta, \Lambda}
\;(\text{Weakening})
\qquad\qquad
\frac{\Gamma \Rightarrow \Theta, A \quad A, \Delta \Rightarrow \Lambda}
     {\Delta, \Gamma \Rightarrow \Theta, \Lambda}
\;(\text{Cut})
\end{equation*}

\noindent\textbf{Operational rules:}
\begin{equation*}
\frac{\Gamma \Rightarrow \Theta, A}
     {\lnot A, \Gamma \Rightarrow \Theta}
\;(\lnot\text{-Left})
\qquad\qquad
\frac{A, \Gamma \Rightarrow \Theta}
     {\Gamma \Rightarrow \Theta, \lnot A}
\;(\lnot\text{-Right})
\end{equation*}
\begin{equation*}
\frac{\Gamma \Rightarrow \Theta, A \quad B, \Gamma \Rightarrow \Theta}
     {A \rightarrow B, \Gamma \Rightarrow \Theta}
\;(\rightarrow\text{-Left})
\qquad
\frac{A, \Gamma \Rightarrow \Theta, B}
     {\Gamma \Rightarrow \Theta, A \rightarrow B}
\;(\rightarrow\text{-Right})
\end{equation*}
\begin{equation*}
\frac{A, \Gamma \Rightarrow \Theta}
     {\wedge\Phi, \Gamma \Rightarrow \Theta}
\;(\wedge\text{-Left, }A \in \Phi)
\qquad\qquad
\frac{\{\Gamma \Rightarrow \Theta, A : A \in \Phi\}}
     {\Gamma \Rightarrow \Theta, \wedge\Phi}
\;(\wedge\text{-Right})
\end{equation*}
\begin{equation*}
\frac{\{A, \Gamma \Rightarrow \Theta : A \in \Phi\}}
     {\vee\Phi, \Gamma \Rightarrow \Theta}
\;(\vee\text{-Left})
\qquad\qquad
\frac{\Gamma \Rightarrow \Theta, A}
     {\Gamma \Rightarrow \Theta, \vee\Phi}
\;(\vee\text{-Right, }A \in \Phi)
\end{equation*}


Here \(\Gamma, \Theta, \Delta, \Lambda, \Phi\) denote finite (possibly empty) sets of formulae; \(A,B\) denote single formulae; and \(\wedge \Phi\) (\(\vee \Phi\)) denotes the conjunction (disjunction) of all formulae in \(\Phi\). Each inference rule consists of one or more sequents above the line, called the \emph{premises}, and a single sequent below the line, called the \emph{conclusion}. The rule states that the conclusion can be derived whenever the premises have been established. The structural rules manage the ``bookkeeping'' of premises, for example by adding formulae or combining derivations. The operational rules govern the logical connectives by decomposing them into their most basic proof-theoretic uses.\footnote{\label{ft:cut}The Cut rule is exceptional in that the formula \(A\) appears in the premises but not in the conclusion. In this sense it literally ``cuts out'' an intermediate formula from the derivation. Gentzen's cut-elimination theorem shows that any derivation using Cut can, in principle, be transformed into one that does not use it. We nevertheless retain Cut here because it makes proofs substantially shorter; see \cite{gb84} for discussion.} On a first reading, the reader need not master each rule in detail; what matters for the analysis below is that the system provides an explicit way of tracking which conclusions follow from which premises.\footnote{Each inference rule actually captures a basic pattern of reasoning. For example, the $\lnot$-Left rule indicates that from $\Gamma \Rightarrow \Theta, A$, one can derive $\lnot A, \Gamma \Rightarrow \Theta$. This rule formalizes the contrapositive relationship in proof theory. The underlying idea is that if we can obtain $A$ from the antecedent $\Gamma$ (possibly along with other conclusions $\Theta$), then adding $\lnot A$ to those same antecedent creates a contradiction; from a contradiction, we can derive anything—so the conclusion $\Theta$ follows immediately. To see it more clearly, consider a simple case where $\Gamma = \{B\}$ and $\Theta = \emptyset$. Then, the premise $B \Rightarrow A$ means that from $B$, we can obtain $A$, and the derived conclusion $\lnot A, B \Rightarrow  \text{ \ \ }$ means that the assumptions $\lnot A$ and $B$ together are contradictory. This makes perfect sense: if $B$ allows us to derive $A$, then assuming both $B$ and $\lnot A$ simultaneously leads to contradiction, because we would have both $A$ and $\lnot A$.}


\paragraph{Logical axiom schema.} All tautologies of the form $A \Rightarrow A$ are axioms, for any formula $A$.\footnote{One distinctive feature of Gentzen-style systems, as compared with Hilbert-style systems \citep[which is used in many introductory textbooks, such as ][]{em15}, is that they minimize axioms and shift the logical burden to inference rules. This is especially useful here, because our interest lies not merely in provability, but in the structure of derivations and in the dependence of conclusions on premises.}

\subsection{Non-logical axioms for TU games}\label{subsec:reasoning}

The logical axioms above are content-free. To reason about a specific TU game, agents also need \emph{non-logical axioms} encoding their basic numerical competence: the ability to recognize when one payoff vector weakly dominates another with respect to a coalition.

\begin{itemize}
  \item $\ \  \ \Rightarrow y \geq_S x$ \quad for all $y \geq_S x \in
    \mathsf{At}_\mathrm{wd}$ with $y_i \geq x_i$ for each $i \in S$;
  \item $\ \  \ \Rightarrow \lnot(y \geq_S x)$ \quad for all $y \geq_S x
    \in \mathsf{At}_\mathrm{wd}$ with $y_i < x_i$ for some $i \in S$.
\end{itemize}

These axioms have empty antecedents: they are unconditional truths that require no coalition-level information to derive. Together they establish that $y \geq_S x$ holds if and only if $y_i \geq x_i$ for all $i \in S$---the minimal numerical reasoning capacity shared by all rational agents. Notice these axioms do \emph{not} provide any information about which payoff vectors are feasible for which coalitions. That information, captured by sentences of the form $y^S$ or $\lnot y^S$, must enter as explicit antecedent---which is precisely the sense in which an agent's information is a measurable, syntactic object in our framework.


%

\section{Information Requirements Underlying the Core}\label{sec:cor}

The example in Section \ref{subsec:example} demonstrated, for a two-player game, how proof-theoretic reasoning tracks information about coalition structure: agent $1$ does not reject a payoff vector when the sequent asserting that her $\textsf{R}_1(x)$ follows from the information she holds is unprovable.
The present section generalizes this to arbitrary TU games and addresses the central question of the paper: the informational requirements underlying the core. The core is conventionally defined in terms of feasible coalitional deviations. Yet what should individual agents know to recognize such deviations, and how must this information be distributed for their decisions to recover the core?


We address these questions by introducing a decision criterion based on justified rejection (Section~\ref{subsec:criterion}). An agent rejects a proposed payoff vector only when she can derive the existence of a feasible coalitional deviation that strictly benefits her without harming the other members. Failure to derive such a deviation does not imply that she can establish its nonexistence; her information may simply be insufficient to reach either conclusion. This distinction allows the framework to accommodate ignorance without identifying it with infeasibility.

We first specify the informational structure supporting these individual decisions (Section \ref{subsec:info}) and establish several basic properties of rejection (Section \ref{subsec:auxiliary}). We then characterize the informational distributions under which the set of unopposed payoff vectors coincides exactly with the core of a fixed TU game (Section \ref{subsec:load-bearing}). This characterization yields a measure of minimum informational burden, defined by the smallest collection of coalition information sufficient to recover the core.



\subsection{The rejection criterion}\label{subsec:criterion}
%

For a general TU game $G = (N, v)$ and agent $i \in N$, let $\mathcal{N}_i := \{S \subseteq N : i \in S\}$ denote the collection of all coalitions containing $i$. Given a proposed payoff vector $x \in \mathsf{PV}(G)$, the \emph{decision criterion} for agent $i$ to \emph{reject} $x$ is the sentence
\begin{equation}\label{eq:core-crit}
  \textsf{R}_i(x) := 
    \bigvee_{S \in \mathcal{N}_i}
    \bigvee_{y \in \mathsf{PV}(G)}
    \Bigl(y^S \wedge (y \geq_S x) \wedge (y >_{\{i\}} x)\Bigr)
  \tag{CORE-CRIT}
\end{equation}

The formula $\textsf{R}_i(x)$ asserts that agent $i$ finds some coalition $S$ with $i \in S$ and some payoff vector $y$ such that $y$ is feasible for $S$, weakly improves every member of $S$ relative to $x$, and strictly improves agent $i$. Comparing with the example in Section \ref{subsec:example}, (\ref{eq:core-crit}) is the natural $n$-player extension of formula~(\ref{exp02:target}): the disjunction now ranges over all coalitions in $\mathcal{N}_i$ rather than just $\{1\}$ and $\{1,2\}$. The criterion expresses the possibility of a profitable coalitional deviation from the perspective of an individual agent.

In our framework, the statement $\vdash \Gamma_i \Rightarrow \textsf{R}_i(x)$ formalizes the idea that, given the information $\Gamma_i$ she possesses, agent $i$ can justify rejecting $x$ according to the decision criterion $\textsf{R}_i(x)$ and would reject $x$.\footnote{Strictly speaking, the statement \(\vdash \Gamma_i \Rightarrow \textsf{R}_i(x)\) is formulated at the meta-level: it is a claim about what is derivable from the information set \(\Gamma_i\) in the proof system \emph{of an agent}. One might instead introduce an additional belief operator such as \(B_i(\cdot)\) and replace the statement as $\vdash B_i(\Gamma_i) \Rightarrow B_i(\textsf{R}_i(x))$. We do not pursue this approach, since our concern is not the interaction between different agents' beliefs, but the inferences an individual agent can draw from the information she possesses. The information set \(\Gamma_i\) already represents her informational state syntactically, and the derivability relation captures her reasoning from that information. Introducing an additional belief operator would therefore add a layer of formalism without contributing to the present analysis. See, e.g., \citet{k02} for a general discussion, and compare \citet{ks02}, where an explicit belief operator is introduced.}
Conversely, $i$ \emph{does not reject} $x$ if and only if $\not\vdash \Gamma_i \Rightarrow \textsf{R}_i(x)$. Here, note that $\not\vdash \Gamma_i \Rightarrow \textsf{R}_i(x)$ includes two cases: (1) she can prove that no such deviation exists, i.e.,  $\vdash \Gamma_i \Rightarrow \lnot \textsf{R}_i(x)$, (2) her information is insufficient to establish either its existence or its absence, i.e., $\not\vdash \Gamma_i \Rightarrow \textsf{R}_i(x)$ and $\not\vdash \Gamma_i \Rightarrow \lnot\textsf{R}_i(x)$.

A payoff vector $x$ is \emph{unopposed} if every $i \in N$ does not reject it.

\subsection{Information structure and decidability}\label{subsec:info}

An agent's information determines which sentences she can use in the antecedent. As argued in Section~\ref{sec:logic}, the only sentences that bear on whether $\textsf{R}_i(x)$ is derivable
are feasibility sentences $y^S$ in  \(\mathsf{At}_\mathrm{fv}(G)\) and their negations $\lnot y^S$: the dominance comparisons $y \geq_S x$ are handled entirely by the non-logical axioms and require no coalition-specific information. Hence, in the following, we focus on the case where the information possessed by $i$, i.e., the antecedent $\Gamma_i$, satisfies,
\begin{equation}\label{eq:info}
\Gamma_i \subseteq  \mathsf{At}_\mathrm{fv}(G) \cup \{\lnot y^S: y^S \in \mathsf{At}_\mathrm{fv}(G)\}
\tag{INFO-FEAS}
\end{equation}


Our first concern is the consistency of an agent's information. Rejection requires a proof of \(\mathsf R_i(x)\), whereas the absence of such a proof suffices for non-rejection. Thus, an agent need not be able to derive either \(\mathsf R_i(x)\) or its negation. What we must exclude, however, is information that supports contradictory conclusions. The following proposition establishes that, under condition (\ref{eq:consis}), it suffices to exclude directly contradictory feasibility literals.

\begin{proposition}[\textbf{Consistency under feasibility information}]
\label{prop:consistency}
Suppose that $\Gamma_i$ satisfying (\ref{eq:info}) and the following consistency condition:
\begin{equation}\label{eq:consis}
\text{For each } y^S \in \mathsf{At}_\mathrm{fv}(G), \text{ if }y^S \in \Gamma_i, \text{ then }\lnot y^S \not\in \Gamma_i
\tag{INFO-CONSIS}
\end{equation}
Then for each $x$, $\not\vdash \Gamma_i \Rightarrow \textsf{R}_i(x) \wedge \lnot\textsf{R}_i(x)$.
\end{proposition}

Condition (\ref{eq:consis}) is nevertheless weak from an economic perspective. It excludes a feasibility sentence and its negation from appearing together, but does not ensure that the agent's information is ``contextually'' consistent with the feasibility structure of the underlying game. For instance, a \(\Gamma_i\) might contain a sentence asserting that a coalition can generate a payoff of 10 while contain another sentence denying that the same coalition can generate 9. We therefore introduce a more structured representation, in which information is organized by coalition. For each coalition about which an agent is informed, she possesses accurate information concerning its feasible and infeasible payoff vectors; for other coalitions, no feasibility information is available to her.

Formally, given $G = (N,v)$, for each $S \subseteq N$, let
\begin{equation*}
\hat{\Gamma}^G(S) : = \{y^S: y^S \in \textsf{At}_{\text{fv}}(G) \text{ and }\sum_{j \in S}y_j \leq v(S)\}
\end{equation*}
and
\begin{equation*}
\Gamma^G(S) :=  \hat{\Gamma}(S) \cup \{\lnot y^S: y^S \in \textsf{At}_{\text{fv}}(G) \text{ and }\sum_{j \in S}y_j > v(S)\}
\end{equation*}
Intuitively, $\Gamma^G(S)$ encodes complete and accurate information about the feasibility of $S$: it affirms exactly the feasible payoff vectors for the coalitions and negates every infeasible vector. In this sense, \(\Gamma^G(S)\) can be regarded as a \emph{syntactic representation} of the information corresponding to $S$.

Given a set of collections $\mathcal{S} \subseteq 2^N$, we define \(\Gamma^G(\mathcal{S}) := \cup_{S \in \mathcal{S}}\Gamma^G(S)\). When \(\mathcal S=\mathcal N_i\) (i.e., all coalitions containing $i$), the agent possesses complete feasibility information about every coalition containing her. When \(\mathcal S=\varnothing\), she possesses no coalition-feasibility information: neither the feasibility nor the infeasibility of any payoff vector is available to her as a premise.\footnote{One can see here that our model shares with the unawareness literature \citep*{hmb13} the idea that agents may lack information about certain cooperative possibilities. The nature of this ignorance differs, however. In unawareness models, agents may fail to conceive of certain states or events entirely, and the formal language available to them is correspondingly impoverished. In our model, every agent commands the full formal language: the sentence $y^S$---``coalition $S$ can achieve $y$''---is grammatically available to every agent regardless of whether she holds it as a premise. What varies across agents is not linguistic competence but informational endowment: an agent who does not know $v(S)$ simply lacks the sentence $y^S$ (or its negation) as a usable premise. This corresponds to a natural and common form of ignorance: one may know that a set of people exists and could in principle cooperate, while remaining uninformed about what their cooperation could achieve. The distinction matters formally because unawareness models generate non-standard logical behavior (an unaware agent cannot assert her own ignorance), whereas our agents, being merely uninformed rather than unaware, reason classically from whatever premises they hold.}

Throughout this section, the underlying TU game \(G\) is fixed, and we suppress its dependence in the notation for information sets and abbreviate \(\Gamma^G(\cdot)\) as \(\Gamma(\cdot)\).

\subsection{Three auxiliary results}\label{subsec:auxiliary}


Before stating the main theorem, we establish three auxiliary results. The first concerns an important asymmetry between core and noncore payoff vectors: whereas rejecting a noncore payoff vector may require information about a profitable coalitional deviation, a core payoff vector cannot be rejected, regardless of the information available to an agent.

\begin{proposition}[\textbf{Core payoff vectors can never be rejected}]
\label{prop: core-payoff}
If $x \in C(G)$, then $\not\vdash \Gamma(\mathcal{S}) \Rightarrow \textsf{R}_i(x)$ for every $i \in N$ and every $\mathcal{S} \subseteq \mathcal{N}_i$.
\end{proposition}

The intuition is direct. If $x$ is in the core, no coalition can profitably deviate from it. Therefore, no matter which coalitions an agent knows about, she will find no feasible deviation: the profitable deviations do not exist, so no information can produce a proof of rejection. 
Core payoff vectors are, in this sense, \emph{informationally robust}: their unopposedness requires no coalitional information at all.

This result also clarifies what the main theorem must accomplish. \autoref{prop: core-payoff} handles the ``if'' direction---core vectors are always unopposed---for free. The entire difficulty lies in the ``only if'' direction: ensuring that \emph{non-core} vectors are rejected by at least one agent. That is where information works, and that is what the condition in \autoref{theorem:load-bearing} is designed to guarantee.

The second observation concerns which information is relevant to the rejection decision.

\begin{proposition}[\textbf{Information about irrelevant coalitions does not affect rejection}]\label{prop:irrelevant}
Let $i \in N$, and $\mathcal{S}, \mathcal{T} \subseteq 2^N$ with $\mathcal{S} \cap \mathcal{T} = \emptyset$ and $\mathcal{T} \cap \mathcal{N}_i = \emptyset$ (that is, no coalition
in $\mathcal{T}$ contains $i$). Then for every $x \in \mathsf{PV}(G)$:
\[
 \not\vdash \Gamma(\mathcal{S}) \Rightarrow \textsf{R}_i(x)
  \quad\Longleftrightarrow\quad
  \not \vdash \Gamma(\mathcal{S} \cup \mathcal{T}) \Rightarrow
  \textsf{R}_i(x).
\]
\end{proposition}

\autoref{prop:irrelevant} shows that adding information about coalitions that do not contain agent $i$ leaves her decision completely unchanged. The reason is transparent from the structure of $\textsf{R}_i(x)$: the criterion ranges only over coalitions in $\mathcal{N}_i$, so only sentences $y^S$ with $S \in \mathcal{N}_i$ (i.e., $i \in S$) can be ``load-bearing'' in a proof of $\textsf{R}_i(x)$ or $\lnot \textsf{R}_i(x)$. Information held by agent $i$ about coalitions she does not belong to is, from a decision-theoretic perspective, entirely wasted here. 


The natural next question is therefore whether there is a sufficient amount of such information that guarantees rejection of every non-core payoff vector. \autoref{prop:full} gives an immediate, though not yet minimal, answer: if every agent knows every coalition that includes her, then unopposed-ness coincides with the core.


\begin{proposition}[\textbf{Full self-relevant information suffices}]\label{prop:full}
For each $x \in \textsf{PV}(G)$ feasible to $N$,
\[
  \vdash \Gamma(\mathcal{N}_i) \Rightarrow \lnot \textsf{R}_i(x) \text{ for all } i \in N \text{ if and only if }x \in C(G) 
\]
\end{proposition}



\autoref{prop:full} establishes more than the absence of rejection. Under full self-relevant information, every agent can positively establish that no profitable deviation is available to her whenever the proposed payoff vector belongs to the core. Thus, non-rejection is no longer a consequence of insufficient information: every core payoff vector is supported by a proof of non-rejectability for each agent. This informational requirement, however, is stronger than necessary for recovering the core. 

\subsection{Load-bearing coalitions and informational characterization of the core}\label{subsec:load-bearing}


\autoref{prop:full} provides an upper bound for informational characterization of the core. This subsection addresses how far we can refine the requirement. 

Fix a TU game $G$. We use $F^S$ to denote the set of payoff vectors in \textsf{PV}$(G)$ that are feasible for $S$. For each vector \(x\), a coalition $S \subseteq N$ \emph{blocks} \(x\) iff there is some $y \in F^S$ with $y$dom$_Sx$. We use $\mathcal{B}(x)$ to denote the set of coalitions that block \(x\). Clearly, for each \(x \in F^N\), \(x\) is a core payoff vector if and only if $\mathcal{B}(x) = \emptyset$. We have the following result.




\begin{proposition}\label{prop:load-bearing}
For each payoff vector \(x\), \(\vdash \Gamma(\mathcal{S}_i) \Rightarrow \textsf R_i(x)\) if and only if \(\mathcal S_i \cap \mathcal{B}(x) \neq \emptyset\).
\end{proposition}

In the sense of \autoref{prop:load-bearing}, each element in $\mathcal{B}(x)$ is called a (potential) \emph{load-bearing} coalition, because information about such a coalition could be effective to reject a non-core payoff vector. 

The following result is the main theorem. It establishes a necessary and sufficient informational condition for the set of unopposed payoff vectors to coincide with the core: for every noncore payoff vector, at least one coalition capable of blocking it must be known to one of its members.


\begin{theorem}[\textbf{Informational characterization of the core}]\label{theorem:load-bearing}
Fix a TU game $G$.
\begin{itemize}
\item[\textbf{(i)}] \textbf{(Sufficiency)} If $\big(\bigcup_{i \in N}\mathcal{S}_i\big) \cap \mathcal{B}(x) \neq
   \emptyset$ for each $x \in F^N\setminus C(G)$, then for each $x \in F^N$:
    \[
  \not\vdash \Gamma(\mathcal{S}_i) \Rightarrow  \textsf{R}_i(x) \text{ for all } i \in N \text{ if and only if }x \in C(G) 
\]  
\item[\textbf{(ii)}] \textbf{(Necessity)} If $\big(\bigcup_{i \in N}\mathcal{S}_i\big) \cap \mathcal{B}(x) =
   \emptyset$ for some $x \in F^N\setminus C(G)$,  then even though $x \notin C(G)$,
    $\not\vdash \Gamma(\mathcal{S}_i) \Rightarrow \textsf{R}_i(x)$ for
    all $i \in N$.
\end{itemize}
\end{theorem}


\autoref{theorem:load-bearing} refines the informational requirement for recovering the core. Rather than requiring information about every coalition, it suffices that the informed coalitions collectively cover the blocking opportunities of every noncore payoff vector. A straightforward implication of \autoref{theorem:load-bearing} is that, when $\big(\bigcup_{i \in N}\mathcal{S}_i\big) = 2^N \setminus \{\emptyset\}$, unopposed-ness characterizes the core. However, \autoref{theorem:load-bearing} allows a smaller collection of coalitions to suffice. Since different coalitions may block the same payoff vector, some coalition information may be redundant. This motivates a game-specific measure of informational burden.

For a TU game \(G\), define its \emph{minimum informational burden} by

\[
\mu(G):=
\min_{\mathcal K\subseteq 2^N\setminus{\{\varnothing\}}}
\bigl\{
|\mathcal K|:
\mathcal K\cap\mathcal B(x)\neq\varnothing
\text{ for every }x\in F^N\setminus C(G)
\bigr\}.
\]
Here, \(\mu(G)\) provides a measure of the coalition-specific informational endowment needed to recover the core collectively.
A collection \(\mathcal K\) satisfying \(\mathcal K\cap\mathcal B(x)\neq\varnothing
\text{ for every }x\in F^N\setminus C(G)\) is called an \emph{informational covering} of \(G\); it is a \emph{minimum informational covering} if \(|\mathcal K|=\mu(G)\). When $\big(\bigcup_{i \in N}\mathcal{S}_i\big) = \mathcal K$ for such a $\mathcal K$, we say that \((\mathcal{S}_i)_{i \in N}\) realizes a \emph{minimum informational covering} of \(G\) for characterizing the core.\footnote{Mathematically, this is a minimum hitting set problem: we seek a family $\mathcal{K}$ of smallest cardinality that intersects every $\mathcal{B}(x)$ for $x\notin C(G)$. Unlike the restricted-blocking literature, which investigates the sufficiency of coalitions satisfying prescribed restrictions (e.g., on their size or composition), our formulation places no such restrictions on the constituent coalitions of $\mathcal{K}$.} Its coalitions provide precisely the information needed, within the present framework, to ensure that every noncore payoff vector encounters a justified rejection. Such a collection need not be unique. The next section extends this characterization to exchange economies.

\section{Coalitional Information Burden in Replica Economies}\label{sec: DS}

\autoref{sec:cor} establishes a general characterization of the coalition-feasibility information required for decentralized individual rejection decisions to recover the core. We now investigate how these informational requirements change as an economy expands. Our central question is whether an increasingly large economy can sustain global coalitional stability through a sparse distribution of information, without requiring each individual to know an ever-growing number of feasible coalitional alternatives.

The replica process underlying the Debreu-Scarf theorem provides a natural setting for this investigation. As an economy is replicated, the number of possible coalitions grows exponentially, while the core converges toward competitive equilibrium. Yet the informational requirements of recovering the core need not grow at the same rate. We show that, in Shapley's (\citeyear{s75}) two-type replica economy, the minimum informational burden grows only linearly with population, and the requisite information can be distributed so that no individual holds more than two coalition-feasibility premises. At the same time, this informational decentralization has an important limit: every sufficient informational covering must contain two almost-grand coalitions of opposite type imbalances. Thus, sparse local information can support global stability, but some economy-wide coalition information is indispensable.

Establishing these results requires extending the proof-theoretic framework beyond finite transferable-utility games. In pure-exchange economies, the space of possible allocations is infinite, so the rejection criterion must accommodate quantification over feasible deviations while preserving finite proofs and informational premises. Section \ref{subsec:setup} introduces the exchange-economy setting and the replica process. Section \ref{subsec:replica} develops the corresponding first-order language, and Section \ref{subsec:count} extends the informational characterization of the core to this setting. Finally, Section \ref{subsec:global} establishes the minimum informational burden in Shapley's replica economy and characterizes both the sufficiency of sparse informational coverings and the necessity of their economy-wide component.

\subsection{Pure exchange economy, NTU game, and $k$-fold replica}\label{subsec:setup}

A \emph{pure-exchange economy} is a tuple
\(
E=(N,D,\{e_i\}_{i\in N},\{u_i\}_{i\in N}),
\)
where \(N\) is the set of agents, \(D\) is the set of commodities, \(e_i\in\mathbb{R}_+^D\) is agent \(i\)'s endowment, and \(u_i:\mathbb{R}_+^D\to\mathbb{R}\) is her utility function. A \emph{(Walrasian) competitive equilibrium} of \(E\) is a pair \((p,x)\), consisting of a price vector \(p\) and an allocation \(x=(x_i)_{i\in N}\), such that:
\begin{itemize}
    \item[(i)] \(\sum_{i\in N} x_i=\sum_{i\in N} e_i\);
    \item[(ii)] \(p\cdot x_i\le p\cdot e_i\) for every \(i\in N\);
    \item[(iii)] for every \(i\in N\) and every \(y_i\in\mathbb{R}_+^D\) with \(p\cdot y_i\le p\cdot e_i\), we have \(u_i(x_i)\ge u_i(y_i)\).
\end{itemize}
An allocation \(x\) is called an \emph{equilibrium allocation} of \(E\) if there exists some price vector \(p\) such that \((p,x)\) is a competitive equilibrium.

From \(E\) we derive the associated \emph{cooperative game without transferable utility} (NTU game)
\(
G(E)=(N,V),
\)
where, for each coalition \(S\subseteq N\),
\[
V(S):=\left\{x=(x_i)_{i\in N}\in (\mathbb{R}_+^D)^N:\sum_{i\in S}x_i\le \sum_{i\in S}e_i\right\}.
\]
An allocation vector $x$ is called \emph{feasible for coalition} $S$ if $x \in V(S)$. As in \autoref{def1}, the core is characterized by efficiency and coalitional stability, but in the NTU setting dominance must be formulated in utility terms. Given \(S\subseteq N\) and allocations \(x,y\), we say that \(y\) \emph{dominates} \(x\) \emph{via} \(S\), written \(y\operatorname{dom}_S x\), if \(u_i(y_i)\ge u_i(x_i)\) for all \(i\in S\), with strict inequality for at least one \(i\in S\).

The \emph{core} of \(G(E)\), denoted \(C(E)\), is the set of all allocations \(x=(x_i)_{i\in N}\in (\mathbb{R}_+^D)^N\) satisfying:
\begin{itemize}
    \item[] \textbf{Efficiency:} \(\sum_{i\in N}x_i=\sum_{i\in N}e_i\);
    \item[] \textbf{Coalitional stability:} there is no coalition \(S\subseteq N\) and no allocation \(y\in V(S)\) such that \(y\operatorname{dom}_S x\).
\end{itemize}

We focus on the simple two-agent, two-commodity pure-exchange economy studied by \cite{s75}, which isolates the core--equilibrium convergence in its cleanest form. The economy \(E\) satisfies:
\begin{itemize}
    \item[\textbf{E1.}] \(e_1=(1,0)\) and \(e_2=(0,1)\);
    \item[\textbf{E2.}] \(u_1=u_2=:u\), where \(u\) is continuous, strictly increasing, strictly concave, and homothetic.
\end{itemize}

Under these assumptions, \(E\) has a unique competitive equilibrium allocation, denoted \(x^E\). The equilibrium allocation belongs to the core, but the core generally contains many other allocations as well. The Debreu--Scarf theorem shows that, as the economy is replicated, the core shrinks and converges to the singleton consisting of the equilibrium allocation.


Formally, for \(k\in\mathbb{N}_+\), the \(k\)\emph{-fold replica} of \(E\) is the economy
\[
E_k=(N_k,D,\{e_j\}_{j\in N_k},\{u_j\}_{j\in N_k}),
\]
where
\(
N_k=\bigcup_{i\in N}\{i_1,\dots,i_k\},
\)
and for each replica \(i_t\) of type \(i\), we set \(e_{i_t}=e_i\) and \(u_{i_t}=u_i\). Thus each original agent \(i\) becomes a \emph{type}, containing \(k\) identical replicas in \(E_k\). 
For an allocation \(x=(x_i)_{i \in N}\) of the original economy, let \(x^{(k)}\) denote its \(k\)-fold replica, i.e., assigning \(x_i\) to every type-\(i\) agent. Define \(C_k:=\{x:x^{(k)}\in C(E_k)\}\).

\medskip
\noindent\textbf{Debreu--Scarf Theorem.} Under the maintained assumptions,
\[
C_{k+1} \subseteq C_k \text{ for each }k \in \mathbb{N}, \text{ and }\bigcap_{k=1}^{\infty}C_k=\{x^E\},
\]
where \(x^E\) is the unique competitive equilibrium allocation of \(E\).

\subsection{Adapting the Language to Replica Economies/Games}\label{subsec:replica}
To extend the framework of Sections ~\ref{sec:logic}--~\ref{sec:cor} to exchange economies, we enrich the propositional language with first-order quantifiers over allocations. This allows the existence of a profitable coalitional deviation to be expressed by a finite formula even when the space of feasible allocations is infinite. The underlying proof-theoretic approach remains unchanged: an agent's information is represented syntactically, and rejection requires the derivability of the corresponding criterion in Gentzen's sequent calculus.

In this section, we fix a pure-exchange economy $E=(N,D, \{e_i\}_{i \in N}, \{u_i\}_{i \in N})$.

\paragraph{Alphabet.} The alphabet consists of:
\begin{itemize}
\item \textbf{Individual variables:} $x,y,z,x^\prime,y^\prime,\ldots$, ranging over allocation vectors in \((\mathbb R_+^D)^N\).
  \item \textbf{Constant symbols:} \(\bar a\) for each \(a\in(\mathbb R_+^D)^N\), where \(\bar a\) names the allocation vector \(a\). When evaluating a particular proposed allocation \(x\), we use a constant symbol \(\bar x\) indicating that proposal.\footnote{Although the language contains uncountably many constant symbols and instances of the axioms, every formula, sequent, and proof remains finite. This is the sense of finitism adopted in the present framework. The countability of individual-variable symbols places no restriction on the cardinality of the domain of quantification or the collection of constant symbols.}
  \item \textbf{Predicate symbols:} $F^S(\cdot)$ and $\succeq_S(\cdot,\cdot)$ for each nonempty coalition $S\subseteq N$. The atomic formula $F^S(\bar a)$ means that the allocation $a$ is feasible for coalition $S$, while $\succeq_S(\bar b,\bar a)$ means that every member of $S$ weakly prefers $b$ to $a$. When no confusion arises, we write $\bar b\succeq_S\bar a$ for $\succeq_S(\bar b,\bar a)$.
  \item \textbf{Logical connectives:} $\lnot$ (not), $\wedge$ (and), $\vee$ (or), $\rightarrow$  (implies).
  \item \textbf{Quantifiers:} \(\forall\) (for all), \(\exists\) (for some).
  \item \textbf{Auxiliary symbols:} parentheses $(\ , )$.
\end{itemize}

\paragraph{Sentences.} Note that in such a predicate logic, an \emph{atomic formula} is obtained by applying predicate symbols to constant symbols. For example, given two vectors in $a,b \in (\mathbb{R}_+^D)^N$, $F^S(\bar a)$ and $\bar b\succeq_S \bar a$ are both atomic formulae. A \emph{formula} is any expression obtained by finitely many applications of:
\begin{itemize}
  \item[\textbf{F0.}] Each atomic formula is a formula;
  \item[\textbf{F1.}] If $A$ is a formula, so is $\lnot A$;
  \item[\textbf{F2.}] If $A$ and $B$ are formulae, so are $A \rightarrow B$, $A \wedge B$, and $A \vee B$.
  \item[\textbf{F3}.] If $A$ is a formula, \(a\) is a constant symbol occurring in $A$, and $y$ is an individual variable not occurring in $A$, then \(\forall y A[y / \bar a]\) and \(\exists y A[y / \bar a]\) are formulae; here \(A[y / \bar a]\) is the result of replacing every occurrence of $\bar a$ in $A$ by $y$.
\end{itemize}
As before, we use $b \succ_S a$ as an abbreviation of the formula $(\bar b \succeq_S \bar a) \wedge \lnot (\bar a \succeq_S \bar b)$.

\paragraph{Rejection criterion with quantification.} The extended language allows us to express the existence of a profitable coalitional deviation using a finite formula, even when the allocation space is infinite. For an allocation $x$, let $\bar x$ be its corresponding constant symbol. The
\emph{rejection criterion} for agent $i$ is the formula
\begin{equation}\label{eq:ext-core-crit}
  \overline{\textsf{R}}_i(\bar x) := 
   \exists y \Bigl(\bigvee_{S \in \mathcal{N}_i}
    \Bigl(F^S(y) \wedge (y \succeq_S \bar x) \wedge (y \succ_{\{i\}} \bar x)\Bigr)\Bigr)
  \tag{EXT-CORE-CRIT}
\end{equation}
The criterion asserts that some coalition containing $i$ has a feasible allocation that weakly benefits every member of the coalition and strictly benefits $i$.

\paragraph{Coalition feasibility and finite information.} A more fundamental modification is required by our commitment to finitism. In Section~\ref{sec:cor}, an agent's information about a coalition
was represented by a finite collection of feasibility sentences. In an exchange economy, however, each coalition has infinitely many feasible allocations, so this representation is no longer available. Yet sequents must have finite antecedents, and an infinite set of premises is not a legitimate proof input. 
We address this by introducing the following non-logical axioms of ``downward closure'' of coalition feasibility . For each payoff vector $a, b$ and non-empty $S \subseteq N$:
\begin{itemize}
\item  $F^S(\bar b) \Rightarrow F^S(\bar a) \text{ whenever }\sum_{j \in S}b_j \geq \sum_{j \in S}a_j$,
\end{itemize}
These axioms may appear to introduce game-specific content into what was previously a purely logical system. Yet the appearance is misleading. The axioms encode not facts about a particular game but a general arithmetic competence: given knowledge of a coalition's endowment bundle, an agent can determine which allocations are achievable. 

Recall that for each $i \in N$, $e_i$ is $i$'s endowment. Define $e:=(e_j)_{j \in N}$. For each collection $S \subseteq N$, define the information about $S$ as \(\Gamma^E(S) := \{F^S(\bar e)\}\). For each \(\mathcal S \subseteq 2^N\), \(\Gamma^E(\mathcal S) := \cup_{S \in \mathcal{S}} \Gamma^E(S)\).
This information set is finite. Moreover, if $S\in\mathcal S_i$, then every allocation $a$
feasible for $S$ satisfies $\sum_{j\in S}a_j\leq\sum_{j\in S}e_j$. Consequently, the feasibility sentence $F^S(\bar a)$ is derivable from $\Gamma^E(\mathcal S_i)$ by the downward closure axiom and Cut.


\paragraph{Utility-based comparison axioms.} The non-logical axioms of Section~\ref{subsec:reasoning} 
expressed dominance in terms of payoff coordinates. In the exchange-economy setting, dominance is determined by 
utility. We replace the earlier axioms with:
\begin{itemize}
  \item $ \ \ \ \Rightarrow \bar b \succeq_S \bar a$ if and only if $u(b_i) \geq u(a_i)$ for each $i \in S$;
  \item $\ \ \ \ \Rightarrow \lnot(\bar b \succeq_S \bar a)$ if and only if $u(b_i) < u(a_i)$ for some $i \in S$.
\end{itemize}
Since all agents in the replica economy share the same utility function $u$ (assumption \textbf{E2}), each agent 
can evaluate dominance for any coalition she belongs to without additional information about others' preferences.\footnote{In a replica economy with heterogeneous utility functions, this axiom would require each agent to know others' utilities---a non-trivial informational assumption. Incorporating it properly would call for modal belief or knowledge operators $B_i(\cdot)$ or $K_i(\cdot)$ \citep[see][]{ks02}. We sidestep this issue here by maintaining assumption \textbf{E2}, which allows us to focus cleanly on the informational burden of coalition feasibility---already 
the central object of interest.}

\paragraph{Additional inference rules.} Four standard quantifier inference rules need to be added to the propositional Gentzen system in Section \ref{sec:logic}:

\noindent\textbf{Universal quantifier:}
\begin{equation*}
\frac{A(\bar a), \Gamma \Rightarrow \Theta}
     {\forall z A(z), \Gamma \Rightarrow \Theta}
\;(\forall\text{-Left})
\qquad\qquad
\frac{\Gamma \Rightarrow \Theta, A(a)}
     {\Gamma \Rightarrow \Theta, \forall z A(z)}
\;(\forall \text{-Right})
\end{equation*}

\noindent\textbf{Existential quantifier:}
\begin{equation*}
\frac{A(a), \Gamma \Rightarrow \Theta}
     {\exists z A(z), \Gamma \Rightarrow \Theta}
\;(\exists \text{-Left})
\qquad\qquad
\frac{\Gamma \Rightarrow \Theta, A(\bar a)}
     {\Gamma \Rightarrow \Theta, \exists z A(z)}
\;(\exists\text{-Right})
\end{equation*}

As before, \(\Gamma\) and \(\Theta\) represent finite collections of formulae. In \((\forall\text{-Left})\) and \((\exists\text{-Right})\), the term \(\bar a\) names a particular allocation: a universal statement can be applied to any particular allocation, while an existential statement can be established by exhibiting one. In \((\forall\text{-Right})\) and \((\exists\text{-Left})\), by contrast, \(a\) denotes a \emph{fresh} parameter (called an \emph{eigenvariable}), not occurring in the conclusion sequent. It represents an arbitrary allocation, ensuring that the inference does not depend on its particular identity.
\subsection{Informational characterization of the core in exchange economies}\label{subsec:count}

It is straightforward to see that the system is consistent. \autoref{prop: core-payoff} is still applicable here: leaving core allocations unopposed does not need coalition information. The problem is still about the information to reject payoff vectors outside the core. \autoref{prop:irrelevant} is also still applicable: to reject a non-core payoff vector, information about irrelevant coalition does not help. 
Previously, Theorem~\ref{theorem:load-bearing} leads to a definition of the minimum informational covering. We first have to extend Theorem~\ref{theorem:load-bearing} here. 

Fix an exchange-economy $E=(N,D,\{e_i\}_{i\in N},\{u_i\}_{i\in N})$. For each $i \in N$, let \(\mathcal S_i \subseteq \mathcal N_i\). Again, for each \(x\), we say that a coalition \(S \subseteq N\) blocks \(x\) iff 
\(y \in V(S)\) and \(y\)dom\(_S x\), and we use \(\mathcal B(x)\) to denote the set of coalitions that blocks \(x\).

The following result extends \autoref{prop:load-bearing} to the exchange-economy setting. Although utility is not transferable, continuity and strict monotonicity ensure that any blocking coalition admits a feasible deviation strictly improving every member. Together with the downward-closure axioms, this preserves the correspondence between coalition information and justified rejection.

\begin{proposition}\label{prop:extTh1}
Let \(E\) satisfy E1–E2. For each \(i\in N\), let \(\mathcal S_i\subseteq\mathcal N_i\). Then, for every proposed allocation $x$, \(\vdash \Gamma^E(\mathcal S_i) \Rightarrow \bar{\textsf{R}}_i(\bar x)\) if and only if some $\mathcal S_i \cap \mathcal B(x) \neq \emptyset$.
\end{proposition}


\autoref{prop:extTh1} establishes that the correspondence between coalition information and individual rejection is preserved in the first-order framework.\footnote{Although its statement closely parallels \autoref{prop:load-bearing}, the proof of \autoref{prop:extTh1} is considerably more involved. In particular, the existential quantifier does not permit the direct extraction of a rejection witness from an arbitrary derivation in classical first-order logic. Establishing the correspondence therefore requires a Herbrand-Theorem-type argument, together with an analysis of the non-logical axioms. See Appendix.} Consequently, the informational characterization of the core in \autoref{theorem:load-bearing} extends to exchange economies, despite the infinite allocation space.
\begin{corollary}[\textbf{Informational characterization of the core in pure-exchange economies}]\label{corollary:load-bearing}
Let \(E\) satisfy \textbf{E1}--\textbf{E2}, and let \(\mathcal S_i\subseteq\mathcal N_i\) for every \(i\in N\).
\begin{itemize}
\item[\textbf{(i)}] \textbf{(Sufficiency)} If $\big(\bigcup_{i \in N}\mathcal{S}_i\big) \cap \mathcal{B}(x) \neq
   \emptyset$ for each $x \in V(N)\setminus C(E)$, then for each $x \in V(N)$:
    \[
  \not\vdash \Gamma^E(\mathcal{S}_i) \Rightarrow  \overline{\textsf{R}}_i(\bar x) \text{ for all } i \in N \text{ if and only if }x \in C(E) 
\]  
\item[\textbf{(ii)}] \textbf{(Necessity)} If $\big(\bigcup_{i \in N}\mathcal{S}_i\big) \cap \mathcal{B}(x) =
   \emptyset$ for some $x \in V(N)\setminus C(E)$,  then even though $x \notin C(E)$,
    $\not\vdash \Gamma^E(\mathcal{S}_i) \Rightarrow \overline{\textsf{R}}_i(\bar x)$ for
    all $i \in N$.
\end{itemize}
\end{corollary}

Analogously to the TU-game setting, we define the \emph{minimum informational burden} of \(E\) as

\[
\mu(E):=\min_{\mathcal K\subseteq 2^N\setminus\{\varnothing\}}
\left\{|\mathcal K|:\mathcal K\cap\mathcal B(x)\neq\varnothing
\text{ for every }x\in V(N)\setminus C(E)\right\}.
\]

A collection \(\mathcal K\) with \(\mathcal K\cap\mathcal B(x)\neq\varnothing
\text{ for every }x\in V(N)\setminus C(E)\) attaining this minimum is called a \emph{minimum informational covering} of \(E\).

\subsection{Sparse information and global market stability}\label{subsec:global}

As an economy expands, the number of possible coalitions grows exponentially with its population. One might therefore expect that recovering coalitional stability through decentralized individual reasoning requires an increasingly large informational endowment. 
The result in this subsection shows that such informational growth is not inevitable, while also identifying an irreducible economy-wide component of the information required for the core. 
In Shapley's (\citeyear{s75}) replica economy, a collection of coalition-feasibility facts growing only linearly with the population suffices to identify every allocation outside the core. Moreover, these facts can be distributed so that no individual needs information about more than two coalitions, regardless of the size of the economy.

The key lies in how information is organized. A network of bilateral coalitions suffices to enforce efficiency and equal treatment among agents of the same type. Once these conditions hold, two additional coalitions, each comprising all but one agent, determine which divisions between the two types are stable. Thus, information about a small collection of local and economy-wide coalitions can jointly support global coalitional stability, without requiring any individual to know the feasibility of a large number of potential coalitional deviations.

To establish this result, consider the \(k\)-fold replica economy \(E_k\) introduced in Section \ref{subsec:setup}. For \(k\geq 1\), index the agents of each type by \(r=1,\ldots,k\), with the convention \(2_{k+1}=2_1\). Define

\[
\mathcal P_k
={}\bigl\{\{1_r,2_r\}:r=1,\ldots,k\bigr\} \cup\bigl\{\{1_r,2_{r+1}\}:r=1,\ldots,k\bigr\}
\]

and let

\[
\mathcal L_k=
\mathcal P_k\cup
\bigl\{
N_k\setminus\{1_k\},
N_k\setminus\{2_k\}
\bigr\}.
\]

Note that \(|\mathcal L_k| = 2k +2\) for \(k \geq 2\). The \(2k\) bilateral coalitions in \(\mathcal P_k\) form a single alternating cycle connecting all agents of the economy. Each agent belongs to exactly two of these coalitions, both with agents of the opposite type. The remaining two coalitions contain \(2k-1\) agents each, differing only in the type of agent excluded. We call a coalition \emph{balanced} if it contains the same number of agents of each type.


The cyclic construction is not quite cardinality-minimal when \(k\geq3\). Nevertheless, the minimum can be attained through a remarkably small modification. Let \(P=\{1_1,2_1\}, Q=\{1_2,2_3\}
\), and define, for \(k\geq3\),
\(
\mathcal H_k
=
\bigl(\mathcal P_k\setminus\{P,Q\}\bigr)
\cup\{P\cup Q\}\), and
\[\mathcal F_k = \mathcal H_k\cup
 \bigl\{N_k\setminus\{1_k\},\;N_k\setminus\{2_k\}\bigr\}.
\]
Note that \(|\mathcal F_k| = |\mathcal L_k| - 1 = 2k+1\). \(\mathcal F_k\) replaces two disjoint bilateral coalitions with a single four-person coalition. All other---in particular the two almost-grand---coalitions remain unchanged.

\begin{figure}[htbp]
    \centering

    \begin{subfigure}[b]{0.46\textwidth}
        \centering
       \includegraphics[width=\linewidth]{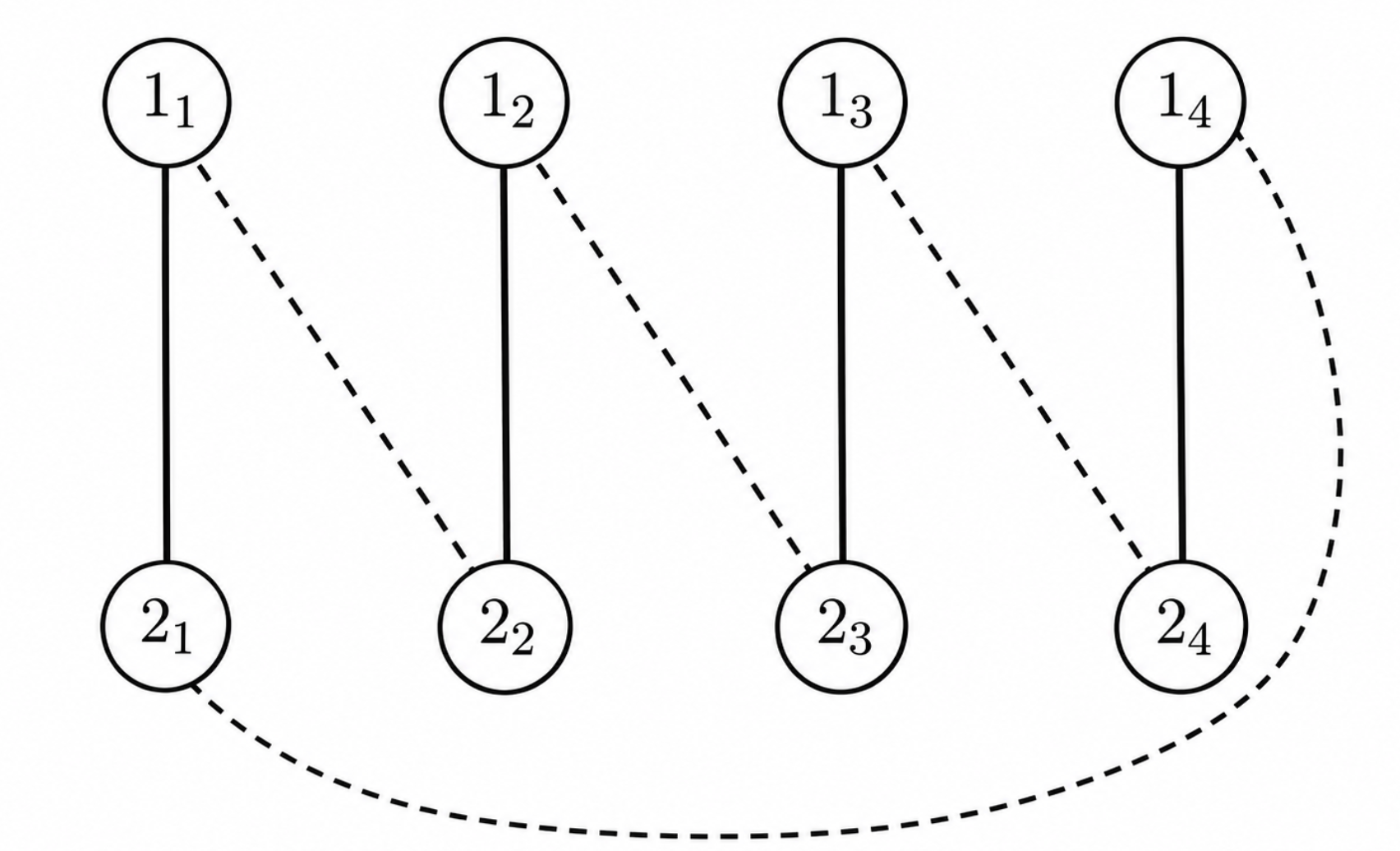}
       \caption{$\mathcal{L}_4$}
       \label{fig:L4}
   \end{subfigure}
   \hfill
    \begin{subfigure}[b]{0.52\textwidth}
        \centering
       \includegraphics[width=\linewidth]{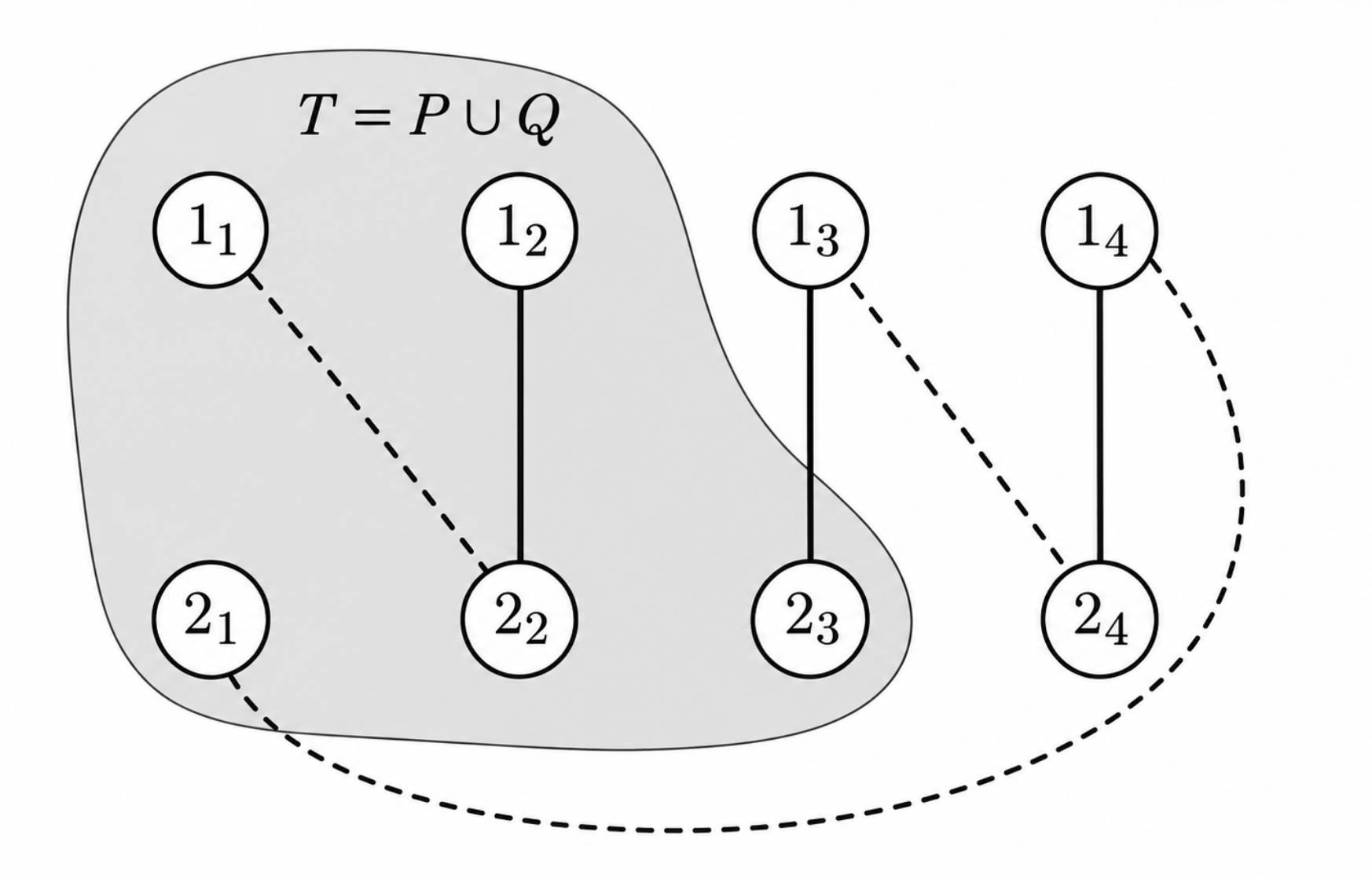}
        \caption{$\mathcal{F}_4$}
       \label{fig:F4}
    \end{subfigure}

   \caption{Two families of load-bearing coalitions for $k=4$.}
    \label{fig:coalitions}
\end{figure}

The following theorem establishes the sufficiency of both constructions, the exact minimum informational burden, and the necessity of certain economy-wide coalitions in every informational covering.

\begin{theorem}[\textbf{Sparse informational foundations of the replica core}]\label{th:info-replica}
 Let \(E_k\) be the \(k\)-fold replica of an economy satisfying \textbf{E1-E2}.
\begin{itemize}
\item[(i)] For every \(k\geq 1\), \(\mathcal L_k\) is an informational covering of \(E_k\),  and for \(k = 1,2\), \(\mathcal L_k\) is a minimum informational covering of \(E_k\), i.e., \(\mu(E_1) = |\mathcal L_1|=3\) and \(\mu(E_2) = |\mathcal L_2|=6\).



\item[(ii)] For every \(k\geq3\), \(\mathcal F_k\) is a minimum informational covering of \(E_k\), and
\(
\mu(E_k)=|\mathcal F_k|=2k+1.
\)

\item[(iii)] Every informational covering of \(E_k\) contains at least one coalition with \(k\) agents of type 1 and \(k-1\) agents of type 2, and at least one coalition with \(k-1\) agents of type 1 and \(k\) agents of type 2.
\end{itemize}

\end{theorem}

Theorem 2 reveals two complementary properties of the informational requirements underlying the replica core: informational scalability and the necessity of an economy-wide informational component. Parts (i) and (ii) establish that the minimum informational burden grows only linearly with population, despite the exponential growth in the number of possible coalitions. More precisely, for \(k\geq3\), we have \(\mu(E_k)=2k+1\), so that the minimum number of distinct coalition-feasibility premises per agent, averaged over the population, satisfies \(
\frac{\mu(E_k)}{|N_k|}
=\frac{2k+1}{2k}
=1+\frac{1}{2k}\longrightarrow1\) as \(k \rightarrow \infty\). 
Moreover, the information can be distributed so that no individual holds more than two coalition-feasibility premises. To see this explicitly, consider first the cyclic covering \(\mathcal L_k\). Assign each bilateral coalition to its type-1 member, and assign the two almost-grand coalitions to agent \(2_1\). Every type-1 agent then holds two premises, while agent \(2_1\) holds the two almost-grand premises. For the minimum covering \(\mathcal F_k\), the same assignment can be retained for the remaining bilateral coalitions; assign the four-person coalition to agent \(1_1\), and again assign the two almost-grand coalitions to agent \(2_1\). No agent holds more than two premises. This individual upper bound is also the smallest possible: since every sufficient covering contains more than \(2k\) distinct coalitions, at least one agent must hold two or more premises.


The structure of the covering is as important as its cardinality. It reflects a division of informational labor between two economically distinct tasks. Balanced coalitions are indispensable for detecting certain departures from equal treatment: as the proof shows, some noncore allocations that treat agents of the same type differently cannot be blocked by any unbalanced coalition. In the simpler construction \(\mathcal L_k\), bilateral coalitions linking agents of opposite types form a connected cycle in which every individual is linked to exactly two agents of the opposite type. Although these relationships are local, their overlapping constraints, together with aggregate feasibility, suffice to ensure that any allocation they leave unblocked satisfies the required equal-treatment condition. In the minimum construction \(\mathcal F_k\), two bilateral coalitions are replaced by a single four-person coalition, preserving this essential role of balanced coalitional information. Thus, a sparse collection of local or small-group informational relationships is sufficient to restrict the potentially stable allocations to the equal-treatment family.

Equal treatment alone, however, does not guarantee coalitional stability. Some allocations that satisfy equal treatment still lie outside the core, and balanced coalitions cannot block them. Detecting these allocations requires unbalanced coalition information. In both constructions, this task is performed by the two almost-grand coalitions, whose nonblocking constraints rule out equal-treatment allocations that lie outside the core. Part (iii) further shows that the role of these coalitions is not merely a feature of our constructions. Every informational covering must contain at least one almost-grand coalition excluding an agent of type \(1\) and another excluding an agent of type \(2\). Thus, while balanced coalition information can be organized through sparse local relationships, some economy-wide coalition information is indispensable for recovering global coalitional stability.


The resulting informational architecture is therefore neither purely local nor centrally omniscient. Most of the requisite information can concern small groups and be distributed across a sparse network of relationships. At the same time, some economy-wide coalition information is indispensable, independently of the particular covering chosen. 
One possible institutional interpretation is that the latter information could be held or certified by a public or coordinating body, while ordinary participants rely primarily on local knowledge. The theorem does not, however, require a government, a central institution, or any particular assignment of the almost-grand coalition premises; yet note that neither does the bounded number of premises imply that their acquisition or processing costs are bounded: although each almost-grand coalition counts as one informational unit in our measure, the scope of the corresponding information expands with the economy.

The constructions exhibit a further form of scalability. Consider the transition from \(E_k\) to \(E_{k+1}\). In the cyclic construction, all existing bilateral relationships are preserved except the closing link \(\{1_k,2_1\}\), which is replaced by \(\{1_k,2_{k+1}\}\) and \(\{1_{k+1},2_1\}\), while the two entrants also form the pair \(\{1_{k+1},2_{k+1}\}\). For \(k\geq3\), the four-person coalition defining \(\mathcal F_k\) can likewise be preserved as the economy expands. Under the informational assignment described above, only two incumbent agents need to update their premises: agent \(1_k\), whose local relationship changes, and agent \(2_1\), who updates the two almost-grand coalition premises. All other incumbent agents can retain their previous informational endowments. Thus, an expanding economy can maintain a sufficient informational covering without requiring a wholesale reorganization of the information possessed by its existing participants.

Together with the classical Debreu-Scarf convergence result, \autoref{th:info-replica} suggests a new perspective on the relationship between market scale and informational requirements. As replication brings the core toward competitive equilibrium, the minimum stock of coalition-feasibility information grows only proportionally with population, and its distribution can keep individual informational stocks uniformly bounded. Yet this decentralization does not eliminate the need for some economy-wide information. In this precise sense, global market stability can rest on sparse individual knowledge together with an irreducible, but limited, global informational component. How such informational structures arise through economic interaction, and whether similar properties hold in more general economies, remain questions for further research.

\section*{Appendix: Proofs}
\begin{proof}[Proof of \autoref{prop:consistency}]
Suppose that the opposite is true, that is, $\vdash \Gamma_i \Rightarrow \textsf{R}_i(x) \wedge \lnot\textsf{R}_i(x)$. By the inference rules for conjunction and negation, together with cut, we obtain $\vdash \Gamma_i \Rightarrow \ \ \ \ $. By Gentzen's Cut-Elimination Theorem, this sequent admits a cut-free derivation. Since every formula in \(\Gamma_i\) is about feasibility, a cut-free proof of \(\Gamma_i\) must starts with instances of the logical axiom schemata and of the non-logical axioms. The logical rules cannot generate a contradiction from a collection of feasibility literals unless some atom occurs both positively and negatively. However, this is excluded by (\ref{eq:consis}). Nor can the proof terminate in a contradiction arising solely from the non-logical axioms, which are assumed consistent and contain no feasibility atoms. Consequently, no cut-free derivation of $\Gamma_i \Rightarrow \ \ \ \ $ exists, a contradiction. Therefore, $\not\vdash \Gamma_i \Rightarrow \textsf{R}_i(x) \wedge \lnot\textsf{R}_i(x)$.
\end{proof}

Proving \autoref{prop: core-payoff} needs the following technical lemma.
\begin{lemma}[\textbf{Rejection witness}]\label{lemma:rej}
Suppose that \(\Gamma_i\) satisfies (\ref{eq:info}) and (\ref{eq:consis}). Then $\vdash \Gamma_i \Rightarrow \textsf{R}_i(x)$ if and only if there exists $S \subseteq \mathcal{N}_i$ and $y \in \textsf{PV}(G)$ such that \(y^S \in \Gamma_i\), $y$ dominates $x$ via $S$, and $y_i$ is strictly larger than $x_i$ (in other words, \( \ \ \ \Rightarrow y \geq_S x\), \(\ \ \ \ \Rightarrow y \geq_{\{i\}} x\), and  \(\ \ \ \ \Rightarrow \lnot(x \geq_{\{i\}} y)\) are all non-logical axioms).
\end{lemma}

\begin{proof}
The if-part is straightforward from inference rules \(\wedge\)-Right, \(\vee\)-Right, and Weakening. Here we show the only-if part. First, define $\Psi_i(x):=\{y^S: S \subseteq \mathcal{N}_i$, $y \in \textsf{PV}(G)$, $y$ dominates $x$ via $S$, and $y_i$ is strictly larger than $x_i\}$. One can easily see that \(\textsf{R}_i(x)\) and \(\Psi_i(x)\) are logically equivalent, i.e., $\vdash \textsf{R}_i(x) \Rightarrow \Psi_i(x)$ and $\vdash  \Psi_i(x) \Rightarrow \textsf{R}_i(x)$. 

Suppose that \(\Gamma_i\) satisfies (\ref{eq:info}) and (\ref{eq:consis}) and $\vdash \Gamma_i \Rightarrow \textsf{R}_i(x)$. By Cut, $\vdash \Gamma_i \Rightarrow \Psi_i(x)$. Let \(\Gamma_i^+ := \{z^T: z^T \in \Gamma_i\}\) and \(\Gamma_i^- := \{z^T: \lnot z^T \in \Gamma_i\}\). By (\ref{eq:consis}), \(\Gamma_i^+ \cap \Gamma_i^- = \emptyset\). Hence, one can see that $\vdash \Gamma_i^+ \Rightarrow \Gamma_i^- \Psi_i(x)$. Note that since every formula in $\Gamma_i^+ \Rightarrow \Gamma_i^- \Psi_i(x)$ is atomic, a Cut-free proof of it should start from instances of the logical axiom schema. Since \(\Gamma_i^+ \) and \(\Gamma_i^-\) are disjoint, such an instance should not contain any element in \(\Gamma_i^-\). Therefore, for some $y^S \in \Psi_i(x)$, \(y^S \in \Gamma_i^+ \subseteq \Gamma_i\). By definition, this follows that $y$ dominates $x$ via $S$, and $y_i$ is strictly larger than $x_i$.
\end{proof}

\begin{proof}[Proof of \autoref{prop: core-payoff}] 
Suppose, by contradiction, that $\vdash \Gamma(\mathcal{S}) \Rightarrow \textsf{R}_i(x)$ for some $i$ and some $\mathcal{S} \subseteq \mathcal{N}_i$. By Lemma \autoref{lemma:rej}, it follows that \(y^S \in \Gamma(\mathcal S)\), $y$ dominates $x$ via $S$, and $y_i$ is strictly larger than $x_i$, which means that $y$ dominates $x$ via $S$, contradicting $x \in C(G)$.
\end{proof}

\begin{proof}[Proof of \autoref{prop:irrelevant}]
 By Lemma \autoref{lemma:rej}, \(\textsf{R}_i(x)\) is derivable if and only if the antecedent contains a feasibility sentence \(y^S\) witnessing a profitable deviation by some coalition $S \in \mathcal{N}_i$. Since $\mathcal{S} \cap \mathcal{T} = \emptyset$, adding information about coalitions in \(\mathcal{T}\) cannot introduce such a witness. Hence the equivalence follows.
 \end{proof}

The proof of \autoref{prop:full} needs a small technical result.
\begin{lemma}[Cancellation of negation]\label{lemma:meta}
If $\text{ \ }\vdash \Gamma \Rightarrow \lnot A$, then $\vdash \Gamma, A \Rightarrow \text{ \ \ \ }$.
\end{lemma}
\begin{proof}
\begin{equation*}
\frac{\frac{\vdots}{\Gamma \Rightarrow \lnot A} \text{ \ \ \ \ } \frac{A \Rightarrow A}{\lnot A, A \rightarrow \text{ \ \ \ }}(\lnot\text{-Left})}{\Gamma, A \Rightarrow \text{ \ \ \ }}(\text{Cut})
\end{equation*}
\end{proof}

\begin{proof}[Proof of \autoref{prop:full}] 
\emph{(If)} Suppose that \(x \in C(G)\). Fix \(i \in N\). For each \(S \in \mathcal{N}_i\) and each $y \in \textsf{PV}(G)$, either $y$ is not feasible for $S$ (that is, \(\lnot y^S \in \Gamma(\mathcal{N}_i)\)), or $y$ does not dominate $x$ via \(S\) (that is, either \(\ \ \ \ \Rightarrow \lnot (y \geq_S x)\) is a non-logical axiom, or \(\ \ \ \ \Rightarrow \lnot (y >_{\{i\}} x)\) is derivable from some non-logical axioms). Applying Lemma \autoref{lemma:meta}, followed by \(\wedge\)-Left and Weakening as needed, yields \( \vdash \Gamma(\mathcal{N}_i), y^S \wedge (y \geq_S x) \wedge (y >_{\{i\}} x) \Rightarrow \ \ \ \ \). Since this holds for each \(S \in \mathcal{N}_i\) and each $y \in \textsf{PV}(G)$, applications of \(\vee\)-Left and \(\lnot\)-Right yield \( \vdash \Gamma(\mathcal{N}_i) \Rightarrow \lnot \mathsf R_i(x)\).

\emph{(Only if)} Suppose that \(x \not\in C(G)\). Then some \(S \subseteq N\) has a feasible payoff vector $y \in \textsf{PV}(G)$ such that $y$ dominates $x$ via $S$ with $y_i > x_i$ for some $i \in S$. Since \(y^S \in \Gamma(\mathcal{N}_i)\), by Weakening \( \vdash \Gamma(\mathcal{N}_i) \Rightarrow y^S\). Since $y$ dominates $x$ via $S$ and $y_i > x_i$, The non-logical axioms yield \( \vdash \ \ \ \ \Rightarrow (y \geq_S x) \wedge (y >_{\{i\}} x)\). Therefore, \( \vdash \Gamma(\mathcal{N}_i) \Rightarrow y^S \wedge (y \geq_S x) \wedge (y >_{\{i\}} x)\). Applying \(\vee\)-Right yields \( \vdash \Gamma(\mathcal{N}_i) \Rightarrow \textsf{R}_i(x)\). By \autoref{prop:consistency}, it follows that \( \not\vdash \Gamma(\mathcal{N}_i) \Rightarrow \lnot\textsf{R}_i(x)\).
\end{proof}

\begin{proof}[Proof of \autoref{prop:load-bearing}]
The only-if part is directly derived from Lemma \autoref{lemma:rej}. For the if-part, let $S \in \mathcal S_i \cap \mathcal{B}(x)$. Then there is some \(y \in F^S\) with \(y\)dom\(_S x\). If \(y_i > x_i\), we keep \(y\); otherwise we choose some \(j \in S\) with \(y_j  >x_j\) and define a vector \(z\) by 
follows:
\begin{equation*}
				z_k = 
				\begin{cases}
				        y_k & \text{if }k \neq i,j, \\
					y_k + 1 & \text{if } k = i, \\
					y_k-1 & \text{if } k = j. \\
				\end{cases}    
\end{equation*}
Since we consider here only natural numbers, it follows that \(z^S \in \Gamma(S) \subseteq \Gamma(\mathcal{S}_i)\). In each case, we have found a vector (the former \(y\), the later \(z\)) which is feasible for \(S\), dominates \(x\), with the \(i\)-th coordinate strictly bigger than \(x_i\). By Lemma \autoref{lemma:rej}, \(\vdash \Gamma(\mathcal{S}_i) \Rightarrow \textsf R_i(x)\).
\end{proof}

\begin{proof}[Proof of \autoref{theorem:load-bearing}]
\textbf{(i)} Suppose $\big(\bigcup_{i \in N}\mathcal{S}_i\big) \cap \mathcal{B}(x) \neq
   \emptyset$ for each $x \in F^N\setminus C(G)$. It directly follows from Lemma \autoref{lemma:rej} that $\vdash \Gamma(\mathcal{S}_{i})\Rightarrow \textsf{R}_{i}(x)$ implies $x \not\in C(G)$. For the other way round, assume that $x \not\in C(G)$. Then, since \(\big(\bigcup_{i \in N}\mathcal{S}_i\big) \cap \mathcal{B}(x) \neq \emptyset\), for some \(i \in N\), \(\mathcal S_i \cup \mathcal{B}(x) \neq \emptyset\). Hence, by \autoref{prop:load-bearing}, \(\vdash \Gamma(\mathcal{S}_i) \Rightarrow \textsf R_i(x)\).

\textbf{(ii)} $\big(\bigcup_{i \in N}\mathcal{S}_i\big) \cap \mathcal{B}(x) = \emptyset$ for some $x \in F^N\setminus C(G)$. It implies that for all $i \in N$, \(\mathcal S_i \cap \mathcal{B}(x) = \emptyset\). By \autoref{prop:load-bearing}, it follows that $\not\vdash \Gamma(\mathcal{S}_i) \Rightarrow \textsf{R}_i(x)$ for all $i \in N$.
\end{proof}

To show this \autoref{prop:extTh1}, we need a lemma. 

\begin{lemma}[\textbf{Restricted existential-witness property}]\label{lemma:Herbrand}
Suppose \(\vdash \Gamma^E(\mathcal S_i) \Rightarrow \overline{\textsf{R}}_i(\bar x)\). Then there exists \(S_i \in \mathcal S_i\) and a allocation \(\bar a\) such that \(\vdash \Gamma^E(\mathcal S_i) \Rightarrow F^S(\bar a)\), and the utility-based comparison axioms establish \(\vdash  \ \ \ \ \Rightarrow \bar a \succeq_S \bar x\) and \(\vdash  \ \ \ \ \Rightarrow \bar a \succ_{\{i\}} \bar x\).
\end{lemma}

\begin{proof}[Proof of Lemma \ref{lemma:Herbrand}]
In the following, we use \(\mathtt{D}\) to denote the set of downward-closure axioms, and \(\mathtt{U}\) the utility-based comparison axioms. Given payoff vector \(x\), define
\[
 \overline{\textsf{R}}^o_i(y, \bar x) := 
  \Bigl(\bigvee_{S \in \mathcal{N}_i}
    \Bigl(F^S(y) \wedge (y \succeq_S \bar x) \wedge (y \succ_{\{i\}} \bar x)\Bigr)\Bigr)
\]
So that \(\overline{\textsf{R}}^o_i(\bar x)= \exists y \overline{\textsf{R}}^o_i(y, \bar x)\). 

Suppose that \(\vdash \Gamma^E(\mathcal S_i) \Rightarrow \overline{\textsf{R}}_i(\bar x)\), that is, \(\vdash \Gamma^E(\mathcal S_i) \Rightarrow \exists y \overline{\textsf{R}}^o_i(y, \bar x)\). Since proofs are finite, only finitely many instances of the non-logical axioms occur in a proof of it. Let \(D \subseteq \mathtt{D}\) and \(U \subseteq \mathtt{U}\) be finite collections containing these instances. We define
\[D^* := \{F^S(\bar b) \rightarrow F^S(\bar a): (F^S(\bar b) \Rightarrow F^S(\bar a)) \in D\}
\]
\[U^* := \{\bar b \succeq_S \bar a: (\ \ \ \ \Rightarrow \bar b \succeq_S \bar a) \in U\} \cup  \{\lnot(\bar b \succeq_S \bar a): (\ \ \ \ \Rightarrow \lnot (\bar b \succeq_S \bar a)) \in U\}
\]
Since both \(D^*\) and \(U^*\) are finite, by weakening, we have
\(\vdash \Gamma^E(\mathcal S_i), D^*, U^* \Rightarrow \exists y \overline{\textsf{R}}^o_i(y, \bar x)\). Note that since all formulae in the antecedent are atomic and quantifier-free. By Herbrand's Theorem \citep[see, e.g., ][p.267]{mgz21}, there are finitely many constant symbols \(\bar a^1, ..., \bar a^m\) such that
\begin{equation}\label{eq:herbrand}
\vdash \Gamma^E(\mathcal S_i), D^*, U^* \Rightarrow \bigvee_{t=1}^{m} \overline{\textsf{R}}^o_i(\bar a^t, \bar x)
\end{equation}
Note that (\ref{eq:herbrand}) has a propositional proof, in which the atoms appearing in (\ref{eq:herbrand}) occur as propositional atoms.

We now explore \(D^*\). We define \(\mathsf H\) as the smallest set of ground feasible atoms satisfying 
\begin{itemize}
\item[(i)] For each \(S \in \mathcal S_i\), \(F^S(\bar e) \in \mathsf H\);
\item[(ii)] If \(F^S(\bar b) \in \mathsf H\), \((F^S(\bar b) \rightarrow F^S(\bar a) \in D^*\), then \(F^S(\bar a) \in \mathsf H\).
\end{itemize}

By definition, the following properties follow immediately:
\begin{itemize}
\item[(1)] If \(F^S(\bar a) \in \mathsf H\), then \(S \in \mathcal S_i\);
\item[(2)] If \(F^S(\bar a) \in \mathsf H\), then \(a\) is feasible for \(S\);

\item[(3)] Every atom in \(\mathsf H\) is derivable from \(\Gamma^E(\mathcal S_i)\) (by applying downward-closures axioms and cut).
\end{itemize}

We now claim that the disjunct in the succedent of (\ref{eq:herbrand}) contains a feasibility atom belonging to \(\mathsf H\). Choose a fresh atom \(p\),\footnote{For example, we can let \(p = F^{\tilde{S}}(\bar d)\), where \(i \not\in \tilde{S}\) and \(\bar d\) be any allocation vector.} and we introduce the abbreviation:
\[
\top : = p \vee \lnot p, \ \ \ \perp:= p \wedge \lnot p
\]
These are ordinary propositional formulae, and it is straightforward to see that \(\vdash \ \ \ \Rightarrow \top\) and \(\vdash \perp \Rightarrow \ \ \ \). We define a uniform propositional substitution \(\sigma\) on the ground atoms occurring in (\ref{eq:herbrand}). For each ground feasible atom \(F^S(\bar a)\), let
\begin{equation*}
				\sigma(F^S(\bar a)) = 
				\begin{cases}
				        \top & \text{if }F^S(\bar a) \in \mathsf H, \\
					\perp & \text{if } F^S(\bar a) \not\in \mathsf H.\\
				\end{cases}    
\end{equation*}
For each ground comparison atom \(\bar a \succeq_S\bar b\), let
\begin{equation*}
				\sigma(\bar a \succeq_S\bar b) = 
				\begin{cases}
				        \top & \text{if } \ \ \ \Rightarrow \bar a \succeq_S\bar b\text{ is a non-logical axiom}, \\
					\perp & \text{if otherwise}.\\
				\end{cases}    
\end{equation*}
This is well-defined because the comparison axioms are assumed to be consistent and complete for the relevant numerical comparisons. We then extend \(\sigma\) recursively to compound formulas in a routine manner: \(\sigma(\lnot A)=\lnot\sigma(A)\), \(\sigma(A\wedge B)=\sigma(A)\wedge\sigma(B)\), \(\sigma(A\vee B)=\sigma(A)\vee\sigma(B)\), and \(\sigma(A\rightarrow B)=\sigma(A)\rightarrow\sigma(B)\). Note that we are not assigning semantic truth values to formulas. We are replacing propositional atoms by two explicitly specified formulas, one provable and the other refutable. 

First, we show the following claim.
\begin{claim}\label{claim:sigma}
For every \(A \in \Gamma^E(\mathcal S_i) \cup D^* \cup U^* \), \(\vdash \ \ \ \Rightarrow \sigma(A)\).
\end{claim}
To show Claim \ref{claim:sigma}, we need to consider three cases.

\textbf{Case 1.} \(A \in \Gamma^E(\mathcal S_i)\). By definition, \(A=F^S(\bar e)\) for some \(S\in\mathcal S_i\). Therefore \(A\in\mathsf H\), and henceforth \(\sigma(A) = \top\). Therefore, \(\vdash \ \ \ \Rightarrow \sigma(A)\).

\textbf{Case 2.} \(A \in D^*\), that is, \(A=F^S(\bar a)\rightarrow F^S(\bar b)\) for some coalition $S$ and some allocations \(bar a\) and \(bar b\). There are two possibilities. One is that \(F^S(\bar a) \in \mathsf H\). Then the closure condition (ii) of \(\mathsf H\) implies that \(F^S(\bar b) \in \mathsf H\). Therefore, \(\sigma(A) = \top \rightarrow \top\), and consequently \(\vdash \ \ \ \Rightarrow \sigma(A)\). The other possibility is \(F^S(\bar a) \not\in \mathsf H\). Then \(\sigma(F^S(\bar a)) = \perp\), and, since \(\sigma(A) = \perp \rightarrow \sigma(F^S(\bar b)\), clearly \(\vdash \ \ \ \Rightarrow \sigma(A)\). 

\textbf{Case 3.} \(A \in U^*\). By construction, \(A\) is either a ground comparison atom or its negation. For the former, by definition \(A = \top\); for the later, \(A = \lnot \perp\). In each case, we have \(\vdash \ \ \ \Rightarrow \sigma(A)\). 

Here we have established Claim \ref{claim:sigma}.\(\heartsuit\)

Note that Propositional LK is closed under uniform substitution: replacing each propositional atom throughout a derivation by a fixed formula yields another valid derivation. Indeed, the substitution preserves the initial sequents and commutes with every propositional inference rule, as follows by induction on the derivation height. Applying \(\sigma\) to the derivation in (\ref{eq:herbrand}), we therefore obtain
\begin{equation}\label{eq:replace}
\vdash\;
\sigma(\Gamma^E(\mathcal S_i)),\sigma(D^*), \sigma(U^*),
\Rightarrow
\sigma\left(\bigvee_{r=1}^{m}\Phi_i(\bar a_r,\bar x)\right).
\end{equation}
Claim \ref{claim:sigma} establishes that all formula in the antecedent of (\ref{eq:replace}) is derivable without premises. Since the antecedent contains finitely many formulas, repeated applications of Cut eliminate them, yielding 
\begin{equation}\label{eq:replace2}
\vdash\;
\ \ \ \ 
\Rightarrow
\sigma\left(\bigvee_{r=1}^{m}\Phi_i(\bar a_r,\bar x)\right).
\end{equation}

Suppose, toward a contradiction, that there are no \(r\in\{1,\ldots,m\}\) and \(S\in\mathcal N_i\) satisfying all three conditions:
\[
\begin{aligned}
&F^S(\bar a_r)\in\textsf H,\\
&\ \ \ \Rightarrow\bar a_r\succeq_S\bar x
\text{ is a comparison axiom},\\
&\ \ \ \Rightarrow\bar a_r\succ_{\{i\}}\bar x
\text{ is derivable from the comparison axioms}.
\end{aligned}
\]
Recall that \( \overline{\textsf{R}}^o_i(\bar a_r, \bar x) := 
  \Bigl(\bigvee_{S \in \mathcal{N}_i}
    \Bigl(F^S(\bar a_r) \wedge (\bar a_r \succeq_S \bar x) \wedge (\bar a_r \succ_{\{i\}} \bar x)\Bigr)\Bigr)\). Under what we have just supposed, every conjunctive disjunct contains at least one factor whose substitution is refutable. To be specific, for each pair \(r, S\), at least one of the following occurs:
\begin{itemize}
\item[(I)] \(F^S(\bar a_r)\notin\mathsf H\), so its substitution is \(\perp\).
\item[(II)] The weak-improvement comparison \(\bar a_r \succeq_S \bar x\) fails, so its substitution is \(\perp\).
\item[(III)] The strict-improvement comparison \(\bar a_r \succ_{\{i\}} \bar x\) fails, so the substituted strict-improvement formula is refutable. 
\end{itemize}    
Using the standard conjunction-left rules, it follows that \(
\vdash\;
\sigma\left(
\bigvee_{r=1}^{m}\Phi_i(\bar a_r,\bar x)
\right)
\Rightarrow \ \ \
\). 
 Combining this with (\ref{eq:replace2}), and apply cut, we obtain \(\vdash \ \ \ \Rightarrow \ \ \ \), which is impossible by the syntactic consistency of propositional LK. Our supposition must therefore be false. Hence, there is a pair \(r,S\) satisfying all three positive conditions displayed above, and therefore coalition \(S\) provides a feasible profitable deviation, as required.
\end{proof}

\begin{proof}[Proof of \autoref{prop:extTh1}]
The if part can be shown in almost the same manner as the if part of \autoref{prop:load-bearing}. To see the only-if part, suppose that
\(
\vdash\Gamma^E(\mathcal S_i)
\Rightarrow \overline{\mathsf R}_i(\bar x)\).
By the Lemma \ref{lemma:Herbrand}, there exist \(S\in\mathcal S_i\) and an allocation \(a\in V_E(S)\) such that \(a\succeq_S x\) and \(a\succ_{\{i\}}x\). Thus \(S\) blocks \(x\), and
\(\mathcal S_i\cap\mathcal B_E(x)\neq\varnothing\).
\end{proof}

To prove \autoref{th:info-replica}, we need some explanations and some lemmas. Note that homotheticity and convexity allow us to choose an \emph{ordinally
equivalent} utility representation $v:\mathbb R_+^2\to\mathbb R_+$ that is continuous, concave, homogeneous of degree one, and normalized by \(v(1,1)=1, v(t,t)=t\quad (t\ge 0)\). Note that strict convexity of preferences implies \emph{strict superadditivity} for nonzero, nonproportional bundles, that is, \( v(p+q)>v(p)+v(q)\) whenever $p,q\ne0$ are not proportional.

For completeness, one can define $v(p)$ as the unique $t\ge0$ for which $p$ is indifferent to $(t,t)$. 
Since $v$ and the original utility \(u\) represent the same preferences, they have exactly the same blocking coalitions.

For a coalition $S\subseteq N$, let $\omega(S)$ be its aggregate
\emph{initial endowment}:
\[
 \omega(S)=\sum_{i\in S}\omega_i=(\ell,m),
\]
where $\ell$ and $m$ are its numbers of role-1 and role-2 agents.

\begin{lemma}[\textbf{Blocking criterion}]\label{lem:block}
For every nonempty coalition $S$ and every feasible allocation $z$,
\begin{equation}\label{eq:block}
 S\text{ blocks }z
 \quad\Longleftrightarrow\quad
 v\bigl(\omega(S)\bigr)>\sum_{i\in S}v(z_i).
\end{equation}
\end{lemma}

\begin{proof}[Proof of Lemma \autoref{lem:block}]
If $S$ blocks $z$ via bundles $y_i$ with
$\sum_{i\in S}y_i=\omega(S)$ and $v(y_i)\geq v(z_i)$ for every $i\in S$,
then homogeneity and concavity imply
\[
 v\bigl(\omega(S)\bigr)
 =v\left(\sum_{i\in S}y_i\right)
 \ge\sum_{i\in S}v(y_i)
 >\sum_{i\in S}v(z_i).
\]
Conversely, if the inequality in \eqref{eq:block} holds, choose
$t_i>v(z_i)/v(\omega(S))$ with $\sum_{i\in S}t_i=1$.
This is possible because the sum of the lower bounds is less than one.
The redistribution $y_i=t_i\omega(S)$ is feasible for $S$ and
satisfies
$v(y_i)=t_i v(\omega(S))>v(z_i)$ for every member.
\end{proof}
Note that Lemma \ref{lem:block} actually implies that \(v(\omega(S))\) is the maximum total utility attainable by coalition \(S\) by redistributing its aggregate initial endowment.\medskip

We will use the following symbols of Shapley's endpoint characterization. Define the equal-treatment Pareto allocations
\begin{equation}\label{eq:zc}
 z(c)=\bigl\langle(c,c)^k,(1-c,1-c)^k\bigr\rangle,
 \qquad 0\le c\le1,
\end{equation}
and let
\begin{equation}\label{eq:endpoints}
 c_k^-:=v(k,k-1)-(k-1),\qquad
 c_k^+:=k-v(k-1,k).
\end{equation}
Also, we let \(N_k\) be the set of agents in the \(k\)-replica economy, i.e., \(N_k =\{1_t: t = 1,...,k\} \cup \{2_t: t = 1,...,k\}\); for each \(i = 1,2\), let \(N_k^i = \{i_t: t = 1,...,k\}\). Finally, recall that \(C(E_k)\) is the set of the core of \(E_k\), i.e., \(E\)'s \(k\)-replica

\begin{lemma}[\textbf{Two almost-grand coalitions and the endpoints of \(C(E_k)\)}]\label{lem:endpoints}
For $k\ge2$,
\begin{equation}\label{eq:interval}
 z(c)\in C(E_k)
 \quad\Longleftrightarrow\quad c_k^-\le c\le c_k^+.
\end{equation}
For allocations of the form \eqref{eq:zc}, the only coalitions needed
to enforce \eqref{eq:interval} are
$N_k\setminus\{2_k\}$ and $N_k\setminus\{1_k\}$.
\end{lemma}

\begin{proof}
It is easy to see that $0<c_k^-<c_k^+<1$: by strictly increasing-ness of \(v\), it is clear that \(v(k, k-1) > v(k-1, k-1) = k-1\) (as we have assumed before) and \(v(k-1,k)< v(k,k) = k\). $(k,k-1)$ and $(k-1,k)$ are nonproportional, so strict superadditivity yields
\[
 v(k,k-1)+v(k-1,k)<v(2k-1,2k-1)=2k-1,
\]
which is equivalent to $c_k^-<c_k^+$.\medskip

Now we show (\ref{eq:interval}). First we show \(\Longrightarrow\) is correct.
The coalition $N_k\setminus\{2_k\}$ has profile $(k,k-1)$
and current total utility $k-1+c$, so it blocks $z(c)$ precisely when
$c<c_k^-$. Why? Because if \(z(c)\) is in the core, it should not be blocked by the coalition $N_k\setminus\{2_k\}$, which, by Lemma \ref{lem:block}, means that \(\sum_{i \in N_k \setminus \{2_k\}} v(z(c)_i) \geq v(k, k-1)\) (recall that \(v(\omega (N_k \setminus \{2_k\})) = v(k, k-1)\)). Note that, as we assumed, \(\sum_{i \in N_k \setminus \{2_k\}} v(z(c)_i) = kc + (k-1) (1-c) = c + k-1\). Therefore, if \(z(c)\) is in the core, \(c + k-1 \geq v(k, k-1)\), i.e., \(c \geq c_k^-\).

Likewise, $N_k\setminus\{1_k\}$ has profile $(k-1,k)$ and blocks precisely when $c>c_k^+$. \medskip

Now we show the converse \(\Longleftarrow\). Given a coalition \(S \subseteq N_k\), we define the \emph{profile} of \(S\) as a pair of integers \((\ell, m)\) such that \((\ell,m)=\left(|S\cap N_k^1|,|S\cap N_k^2|\right)\). It is clear that for two coalitions \(S, T\) with the same profile, \(v(\omega(S)) = v(\omega(T))\). Hence we can just use \(v(\ell,m)\) to denote \(v(\omega(S))\). Take any coalition (say, \(S\)), with profile $(\ell,m)$.
By Lemma~\ref{lem:block}, it blocks $z(c)$ if and only if
\begin{equation}\label{eq:profile}
 v(\ell,m)> \sum_{i \in S}z(c)_i = \ell c+m(1-c)=m+(\ell-m)c.
\end{equation}
We now need to discuss the order between \(\ell,m)\). We separate the discussion into several cases. 

Case 1: $\ell=m$. Suppose $\ell=m$. Then \(v(\ell,m) =v(\ell, \ell) = \ell v(1,1) = \ell\); on the other hand, \(\sum_{i \in S}z(c)_i = \ell c + \ell (1-c) = \ell\), i.e., \(v(\ell,m)=\sum_{i \in S}z(c)_i\), contradiction. Therefore, if such a coalition blocks \(z(c)\), it is impossible that $\ell=m$.


Case 2: $\ell>m>0$. Suppose that $\ell>m>0$. Put
$\phi(t)=v(t,1)$. Concavity implies that the secant slope
\[
 q_+(t):=\frac{\phi(t)-\phi(1)}{t-1},\qquad t>1,
\]
is nonincreasing in $t$.\footnote{To see this, suppose that \(1<s<t\). Then, with \(\lambda = \frac{s-1}{t-1}\) (clearly \(\lambda \in (0,1)\)), one can see that \(s = \lambda t + (1-\lambda)1\). Concavity means that \(\phi(s)\geq\lambda\phi(t)+(1-\lambda)\phi(1)\). Subtracting \(\phi(1)\) from both sides, we have \(
\phi(s)-\phi(1) \geq \lambda[\phi(t)-\phi(1)]\), i.e., \(
\phi(s)-\phi(1)
\geq
\frac{s-1}{t-1}[\phi(t)-\phi(1)].
\). Since \(s-1>0\), dividing by \(s-1\) gives \(
\frac{\phi(s)-\phi(1)}{s-1}
\geq
\frac{\phi(t)-\phi(1)}{t-1}\), which implies that \(q_+(s) \geq q_+(t)\) whenever \(1<s<t\).}
 Since $\ell/m\ge k/(k-1)$,
we have \(
 q_+(\ell/m)\le q_+\bigl(k/(k-1)\bigr)=c_k^-
\), and hence
\[
 v(\ell,m)
 =m+(\ell-m)q_+(\ell/m)
 \le m+(\ell-m)c_k^-
 \le m+(\ell-m)c =  \sum_{i \in S}z(c)_i 
\]
whenever $c\ge c_k^-$. Therefore, it is still impossible that such a coalition blocks \(z(c)\).

Case 3: $m=0$. Suppose $m=0$. Then superadditivity gives
\[
 v(k,k-1)\ge v(1,0)+v(k-1,k-1)
 =v(1,0)+k-1,
\]
which follows that \(v(1,0) \leq c_k^-\). So $v(\ell,0)=\ell v(1,0)\le\ell c_k^-\le\ell c = \sum_{i \in S}z(c)_i$. Again, such a coalition \(S\) cannot block \(z(c)\).

Case 4: \(\ell < m\). Suppose that $\ell<m$. Set $t=\ell/m\in[0,1)$ and define
\(
 q_-(t):=\frac{\phi(1)-\phi(t)}{1-t}
\). Note that \(q_-(t)\) is nonincreasing in $t$ by concavity. As $t\le(k-1)/k$, we get
\(
 q_-(t)\ge q_-\bigl((k-1)/k\bigr)=c_k^+\).
Therefore, whenever $c\le c_k^+$,
\[
 v(\ell,m)=m-(m-\ell)q_-(t)
 \le m-(m-\ell)c_k^+
 \le m-(m-\ell)c.
\]
Thus no coalition can block if $c\in[c_k^-,c_k^+]$.
\end{proof}

\begin{proof}[Proof of \autoref{th:info-replica}]
\textbf{(i)} \textbf{We first show that for \(k \geq 1\), \(\mathcal L_k\) is an informational covering of \(E_k\)}.  For \(k = 1\), \(\mathcal L_k = 2^{N_1} \setminus \{\emptyset\}\), so it is automatically an informational covering of \(E_1\). We then show that for \(k \geq 2\), \(\mathcal L_k\) is an informational covering of \(E_k\), that is, for every $k\ge2$ and every \(N_k\)-feasible allocation $z$, \(z\) is in the core of \(E_k\) if and only if no coalition in $\mathcal L_k$ blocks $z$. The only-if part follows from the definition of the core. For the if part,  suppose that no coalition in $\mathcal L_k$ blocks
$z$. Let
\(
 A_r=v(z_{1_r})\) and \(B_s=v(z_{2_s})\). Since every pair $\{1_r,2_r\}$ is unblocked, Lemma~\ref{lem:block} gives $A_r+B_r\ge v(\omega\{1_r, 2_r\}) = 1$. Summing and using feasibility,
\[
 k\le\sum_{r=1}^k(A_r+B_r)
 =\sum_{i\in N_k}v(z_i)
 \le v\left(\sum_{i\in N_k}z_i\right)
 =v(k,k)=k.
\]
All inequalities are equalities, in particular, \(\sum_{i\in N_k}v(z_i)=v\left(\sum_{i\in N_k}z_i\right)\). By strict superadditivity of $v$
for nonproportional nonzero bundles, this equality implies that all nonzero bundles $z_i$ lie
on a common ray. Their sum is $(k,k)$, so every $z_i$ is diagonal:
\[
 z_{1_r}=(A_r,A_r),\qquad z_{2_s}=(B_s,B_s).
\]
Furthermore $A_r+B_r=1$ for all $r$.

For the second collection of pairs \(\{1_r, 2_{r+1}\}\), nonblocking gives $A_r+B_{r+1}\ge1$. Their $k$ left-hand sides also sum to $k$, so each is an equality. Therefore
\[
 A_r+B_r=A_r+B_{r+1}=1
 \quad\Longrightarrow\quad B_r=B_{r+1}.
\]
The cyclic indexing implies $B_1=\cdots=B_k$, and then
$A_1=\cdots=A_k$. Let $A_1=\cdots=A_k =c$ (i.e., $B_1=\cdots=B_k=1-c$. Hence $z=z(c)$.
Also, note that neither almost-grand coalition in \(\mathcal L_k\) blocks \(z(c)\). Since \(z(c)\) is not blocked by \(N_k \setminus \{1_k\}\), by Lemma \ref{lem:block}, it follows that \((k-1)c + k (1-c) \geq v(k-1,k)\), that is, \(c \le k - v(k-1,k)= c_k^+\). In a similar manner, from the fact that \(z(c)\) is not blocked by \(N_k \setminus \{2_k\}\), we can derive that \(c \geq c_k^-\). Therefore, Lemma~\ref{lem:endpoints} gives $c_k^-\le c\le c_k^+$,
and thus $z$ is in the core of \(E_k\). \medskip

\textbf{Now we show that for \(k = 1,2\), \(\mathcal L_k\) is a minimal informational covering of \(E_k\).} First, consider the case \(k=1\). It is straightforward to verify that \(\mathcal L_1=\{N_1,\{1_1\},\{2_1\}\}=2^{N_1}\setminus\{\emptyset\}\) is a cardinality-minimal informational covering of \(E_1\). Indeed, an equal-treatment Pareto allocation \(z(c)\) with \(c<c_1^-\) can be blocked only by \(\{1_1\}\), while one with \(c>c_1^+\) can be blocked only by \(\{2_1\}\). Moreover, any feasible allocation \((z_1,z_2)\) that is individually rational but not of the form \(z(c)\) is Pareto inefficient, and hence can be blocked only by \(N_1\). Thus, each of the three coalitions is indispensable, establishing the minimality of \(\mathcal L_1\).


Then we show that \(\mathcal L_2\) is a minimal informational covering of \(E_2\). We have to show that every informational covering contains at least six coalitions. Recall that the core consists of the equal-treatment allocations
\(z(c)=\big\langle(c,c)^2,(1-c,1-c)^2\big\rangle\) for all \(c\in[c_2^-,c_2^+]\), where \(
c_2^-=v(2,1)-1\) and \(
c_2^+=2-v(1,2)\).\medskip

\noindent
\textit{Step 1. At least two unbalanced coalitions
are necessary.} Recall that, by Lemma \ref{lem:block}, a coalition with profile \((\ell,m)\) \textbf{does not} block the equal-treatment allocation \(z(c)\) if and only if \(v(\ell,m)\leq \ell c+m(1-c)\). Rearranging, we obtain
\(
\ell c+m(1-c)-v(\ell,m)
\geq 0\). Therefore, for a coalition with profile $(\ell,m)$, we define its \emph{nonblocking slack} as
\(
 G_{\ell,m}(c):=m+(\ell-m)c-v(\ell,m)\). For any coalition with profile $(\ell,m)$, \(
 G_{\ell,m}(c) \geq 0\) if and only if the coalition cannot block \(z(c)\). If $\ell=m$, this slack is identically zero. Thus, balanced coalitions\footnote{We call a coalition \(S\) \emph{balanced} whenever \(|S \cap N_k^1| = |S \cap N_k^2| \).} cannot block any
equal-treatment Pareto allocation.

Consider $c<c_2^-$. Since $z(c_2^-)$ belongs to
the core,
\(
G_{\ell,m}(c_2^-)\geq0
\)
for every coalition profile $(\ell,m)$.
If $\ell<m$, then $G_{\ell,m}$ is decreasing
in $c$, and hence
\(
G_{\ell,m}(c)>G_{\ell,m}(c_2^-)\geq0.
\)
Therefore, an allocation $z(c)$ with $c<c_2^-$
can only be blocked by a coalition with
$\ell>m$. Analogously, an allocation $z(c)$ with
$c>c_2^+$ can only be blocked by a coalition
with $\ell<m$. Consequently, every informational covering of \(E_k\) must contain at least one coalition of each type, and hence at least two unbalanced coalitions.\medskip

\noindent
\textit{Step 2. All four bilateral coalitions
are necessary.} The nonempty balanced coalitions in the
two-replica economy are \(\{1_r,2_s\} (r,s\in\{1,2\})\) and the grand coalition $N_2$. We show that every bilateral coalition is indispensable. Suppose, without loss of generality, that a proposed sufficient family
$\mathcal C$ does not contain
\(
S^*=\{1_1,2_1\}.
\) Choose
\(
c_0\in(c_2^-,c_2^+)
\). For $\varepsilon>0$, consider the allocation
$z^\varepsilon$ defined by
\[
\begin{aligned}
z_{1_1}^\varepsilon
  &=(c_0-\varepsilon,c_0-\varepsilon),\\
z_{1_2}^\varepsilon
  &=(c_0+\varepsilon,c_0+\varepsilon),\\
z_{2_1}^\varepsilon
  &=(1-c_0-\varepsilon,1-c_0-\varepsilon),\\
z_{2_2}^\varepsilon
  &=(1-c_0+\varepsilon,1-c_0+\varepsilon).
\end{aligned}
\]
For sufficiently small $\varepsilon>0$,
this allocation is nonnegative. Moreover,
its aggregate consumption equals $(2,2)$,
so it is \(N_2\)-feasible and it cannot be blocked by the grand coalition. Since $v(t,t)=t$, the coalition $S^*$ can block $z^\varepsilon$, because
\(
v(z_{1_1}^\varepsilon)
+v(z_{2_1}^\varepsilon)
=1-2\varepsilon <v(1,1)=1.
\) Thus $z^\varepsilon$ lies outside the core. However, none of the remaining three
bilateral coalitions can block it. Indeed,
\[
\begin{aligned}
v(z_{1_1}^\varepsilon)
+v(z_{2_2}^\varepsilon)&=1,\\
v(z_{1_2}^\varepsilon)
+v(z_{2_1}^\varepsilon)&=1,\\
v(z_{1_2}^\varepsilon)
+v(z_{2_2}^\varepsilon)&=1+2\varepsilon.
\end{aligned}
\]
Each of these coalitions has aggregate
endowment $(1,1)$, whose utility is $1$.


It remains to verify that no unbalanced coalition can block $z^\varepsilon$ when $\varepsilon$ is sufficiently small. Since $c_0$ lies strictly inside the core interval, every unbalanced coalition
has strictly positive nonblocking slack at $z(c_0)$. Indeed, for $\ell\neq m$, the function $G_{\ell,m}(c)$ is affine with nonzero slope $\ell-m$ and satisfies
\(
G_{\ell,m}(c_2^-)\geq0\) and \(G_{\ell,m}(c_2^+)\geq0\). It follows that
\(
G_{\ell,m}(c_0)>0\text{ whenever }\ell\neq m\). As there are finitely many unbalanced
coalitions, we may define
\[
\eta:=
\min_{\substack{\varnothing\neq S\subseteq N_2\\
                 \ell(S)\neq m(S)}}
G_{\ell(S),m(S)}(c_0)>0.
\]

Let
\(
\delta_{1_1}=\delta_{2_1}=-1,
\delta_{1_2}=\delta_{2_2}=1
\). For any unbalanced coalition $S$, its
nonblocking slack at $z^\varepsilon$ is
\[
\begin{aligned}
G_S(z^\varepsilon)
&:=\sum_{i\in S}v(z_i^\varepsilon)
  -v(\omega(S))\\
&=G_{\ell(S),m(S)}(c_0)
  +\varepsilon\sum_{i\in S}\delta_i\\
&\geq\eta-4\varepsilon.
\end{aligned}
\]
Choosing \(\epsilon\) satisfying \(0<\varepsilon<\min\left\{c_0,1-c_0,\frac{\eta}{4}\right\}
\) ensures that no unbalanced coalition can block such a $z^\varepsilon$.

We have thus constructed a non-core allocation that cannot be blocked by any
coalition other than $S^*$. Hence $S^*$ must belong to every sufficient family. The same argument applies to the other three bilateral coalitions by relabeling agents within each role.
Hence all four bilateral coalitions are indispensable.\medskip

\noindent
\textit{Step 3. Conclusion.} By Step 1, every sufficient family must contain at least two unbalanced coalitions. By Step 2, it must also contain all four bilateral coalitions, which are balanced. Therefore, every sufficient family contains at least \(4+2=6\) coalitions. Since $\mathcal L_2$ contains exactly
six coalitions and is sufficient, it is a minimum informational covering of \(E_2\). \medskip

\textbf{(ii)} \textbf{We first show that for \(k \geq 3\), \(\mathcal F_k\) is an informational covering of \(E_k\),} that is, for each \(N_k\)-feasible allocation \(z\), \(z\) is in the core of \(E_k\) if and only if \(z\) is not blocked by any coalition in \(\mathcal F_k\). Still, the only-if part is straightforward from definition. For the if part, Suppose no coalition in $\mathcal F_k$ blocks $z$, and write
$A_r=v(z_{1_r})$, $B_s=v(z_{2_s})$.

Every agent occurs in exactly two coalitions of $\mathcal H_k$:
this is true in the original cycle $\mathcal E_k$, and removing
two disjoint pairs and inserting their union preserves the incidence
count. Each remaining pair has endowment $(1,1)$ and $T$ has
endowment $(2,2)$. Summing their nonblocking inequalities gives
\[
 2\sum_{i\in N_k}v(z_i)
 \ge (2k-2)v(1,1)+v(2,2)
 =2k.
\]
The factor \(2\) appears because each agent belongs to exactly two coalitions in \(\mathcal H_k\).\footnote{For every bilateral coalition \(\{i,j\}\in\mathcal H_k\), we have \(v(z_i)+v(z_j)\geq v(1,1)=1\); for the four-agent coalition \(T\), we have\(\sum_{i\in T}v(z_i)\geq v(2,2)=2\). Summing these inequalities over all coalitions in \(\mathcal H_k\), we obtain \(\sum_{S\in\mathcal H_k}\sum_{i\in S}v(z_i)\geq(2k-2)\cdot1+2\). But the left-hand side can be rearranged as
\[
\begin{aligned}
\sum_{S\in\mathcal H_k}\sum_{i\in S}v(z_i)
&=\sum_{i\in N_k}
\underbrace{|\{S\in\mathcal H_k:i\in S\}|}_{=2}
v(z_i)\\
&=2\sum_{i\in N_k}v(z_i).
\end{aligned}
\]
Therefore, \(2\sum_{i\in N_k}v(z_i)
\geq(2k-2)v(1,1)+v(2,2)=2k\).} Concavity and feasibility give $\sum_i v(z_i)\le v(k,k)=k$.
Hence equality holds throughout, including in every one of the
individual nonblocking inequalities. Strict superadditivity again
implies that all consumption bundles are diagonal.

After deleting $P$ and $Q$ from the alternating cycle, the remaining
pairs form two paths. One path has vertices $1_1,2_2,1_2$; the other
has vertices $2_3,1_3,2_4,1_4,\ldots,2_k,1_k,2_1$
(with the evident shortening when $k=3$).
Equality of the nonblocking constraints along these paths yields
\[
 \begin{aligned}
 A_1=A_2&=c, & B_2&=1-c,\\
 A_3=\cdots=A_k&=d, & B_1=B_3=\cdots=B_k&=1-d.
 \end{aligned}
\]
The equality for $T$ is
\[
 A_1+A_2+B_1+B_3=v(2,2)=2,
\]
so $2c+2(1-d)=2$, whence $c=d$.
Thus $z=z(c)$. Finally, nonblocking by the two almost-grand
coalitions and Lemma~\ref{lem:endpoints} imply that
$z$ is in the core of \(E_k\).\medskip

Fix \(k \geq 3\). Finally, \textbf{we show that any informational covering of \(E_k\) must contain at least two
unbalanced and at least $2k-1$ balanced coalitions.}

\smallskip\noindent
\emph{Step 1: At least two unbalanced coalitions.}
By Lemma~\ref{lem:endpoints}, the core contains precisely the
allocations $z(c)$ with $c\in[c_k^-,c_k^+]$, where
$0<c_k^-<c_k^+<1$. A balanced coalition with profile $(\ell,\ell)$
never blocks $z(c)$ because
\[
 v(\ell,\ell)=\ell=\ell c+\ell(1-c).
\]
Recall that, 
for a coalition with profile $(\ell,m)$, we define its \emph{nonblocking slack} as \(
 G_{\ell,m}(c)=m+(\ell-m)c-v(\ell,m)\), i.e., \(S\) with profile \((\ell, m)\) cannot block \(z(c)\) if and only if \(G_{\ell,m}(c) \geq 0\). 
By the core characterization, $G_{\ell,m}(c_k^-)\ge0$
and $G_{\ell,m}(c_k^+)\ge0$, because both \(c_k^-\) and \(c_k^+\) are boundaries for the core and should not be blocked by any coalition. If $c<c_k^-$, then
$G_{\ell,m}(c)\ge G_{\ell,m}(c_k^-)\ge0$ whenever $\ell\le m$.
Therefore an allocation $z(c)$ with $c<c_k^-$ can be detected
only by a coalition with $\ell>m$. Symmetrically, an allocation
$z(c)$ with $c>c_k^+$ can be detected only by a coalition with
$\ell<m$. Any informational covering must contain at least one
coalition of each type, hence at least two unbalanced coalitions.

\smallskip\noindent
\emph{Step 2: At least $2k-1$ balanced coalitions.}
Choose $c_0\in(c_k^-,c_k^+)$. For any unbalanced coalition,
$G_{\ell,m}$ is affine with nonzero slope and nonnegative at both
endpoints. Consequently,
\begin{equation}\label{eq:slack}
 G_{\ell,m}(c_0)>0\qquad(\ell\ne m).
\end{equation}
Thus every unbalanced coalition has \emph{strictly positive}
nonblocking slack at $z(c_0)$.

Consider the vector space
\[
 \Delta =\left\{\delta\in\mathbb R^{2k}:
 \sum_{r=1}^k\delta_{1_r}=0,\quad
 \sum_{s=1}^k\delta_{2_s}=0\right\}.
\]
It has dimension dim$\Delta := d=2k-2$. For $\delta\in \Delta$, perturb $z(c_0)$
by assigning agent $i$ the bundle
\begin{equation}\label{eq:perturb}
 z_i^\varepsilon=(t_i^0+\varepsilon\delta_i,
                      t_i^0+\varepsilon\delta_i),
 \qquad
 t_i^0=\begin{cases}
 c_0,&i\text{ has role 1},\\
 1-c_0,&i\text{ has role 2}.
 \end{cases}
\end{equation}
For sufficiently small $\varepsilon>0$, the allocation is
nonnegative and \(N_k\)-feasible. If a balanced coalition $S$ has profile
$(\ell,\ell)$, then
\[
 \sum_{i\in S}v(z_i^\varepsilon)-v(\ell,\ell)
 =\varepsilon\sum_{i\in S}\delta_i.
\]
Thus nonblocking by $S$ amounts to the homogeneous linear
inequality
\begin{equation}\label{eq:linear}
 L_S(\delta):=\sum_{i\in S}\delta_i\ge0.
\end{equation}

Suppose $\mathcal C$ contains at most $d$ balanced coalitions,
$S_1,\ldots,S_r$, where $r\le d (=2k-2)$. There exists a nonzero
$\delta\in \Delta$ satisfying all inequalities
$L_{S_j}(\delta)\ge0$, \(j = 1,...,r\). Indeed, consider the linear map
\[
 L:\Delta \longrightarrow\mathbb R^r,\qquad
 L(\delta)=(L_{S_1}(\delta),\ldots,L_{S_r}(\delta)).
\]
If $L$ has a nontrivial kernel, take a nonzero vector in it.
If its kernel is trivial, $r\le\dim \Delta=d$ forces $r=d$ and
$L$ to be an isomorphism; take the inverse image of
$(1,\ldots,1)$. Either way, the desired $\delta\ne0$ exists.

Fix such a $\delta$ and take $\varepsilon>0$ sufficiently small.
By \eqref{eq:linear}, none of the balanced coalitions in
$\mathcal C$ blocks $z^\varepsilon$. 


We next show that, for sufficiently small $\varepsilon>0$, no unbalanced coalition can block $z^\varepsilon$.

Recall that $c_0\in(c_k^-,c_k^+)$ and that, for every
coalition profile $(\ell,m)$, the nonblocking slack is
defined by
\[
G_{\ell,m}(c)
:=\ell c+m(1-c)-v(\ell,m).
\]
By Lemma~\ref{lem:endpoints}, both $z(c_k^-)$ and $z(c_k^+)$ belong
to the core. Hence
\[
G_{\ell,m}(c_k^-)\geq 0,
\qquad
G_{\ell,m}(c_k^+)\geq 0
\]
for every coalition profile $(\ell,m)$.

Notice that $G_{\ell,m}$ is affine in $c$, with slope
$\ell-m$. Thus, whenever $\ell\neq m$, its slope is
nonzero, and the two endpoint values cannot both
vanish. Since $c_0$ lies strictly between the two
endpoints, it follows that
\[
G_{\ell,m}(c_0)>0
\qquad\text{for every }\ell\neq m.
\]

Now consider any unbalanced coalition $S$, with
profile $(\ell,m)$. Under the perturbed allocation
$z^\varepsilon$, its nonblocking slack is
\[
\begin{aligned}
G_S(z^\varepsilon)
&:=\sum_{i\in S}v(z_i^\varepsilon)
   -v(\omega(S))\\
&=\ell c_0+m(1-c_0)
  +\varepsilon\sum_{i\in S}\delta_i
  -v(\ell,m)\\
&=G_{\ell,m}(c_0)
  +\varepsilon L_S(\delta),
\end{aligned}
\]
where
\[
L_S(\delta):=\sum_{i\in S}\delta_i.
\]

Since the economy contains finitely many coalitions,
we may define
\[
\eta:=
\min_{\substack{\varnothing\neq S\subseteq N_k\\
                 \ell(S)\neq m(S)}}
G_{\ell(S),m(S)}(c_0)>0.
\]
Moreover, since $\delta\neq 0$, let
\[
M:=\sum_{i\in N_k}|\delta_i|>0.
\]
For every coalition $S$, we have
\[
|L_S(\delta)|
=\left|\sum_{i\in S}\delta_i\right| \leq \sum_{i\in N_k}|\delta_i|
= M.
\]

Choose $\varepsilon>0$ sufficiently small that
\[
\varepsilon<\frac{\eta}{M}.
\]
Then, for every unbalanced coalition $S$,
\[
\begin{aligned}
G_S(z^\varepsilon)
&=G_{\ell(S),m(S)}(c_0)
  +\varepsilon L_S(\delta)\\
&\geq\eta-\varepsilon M\\
&>0.
\end{aligned}
\]
Therefore, no unbalanced coalition can block
$z^\varepsilon$.

Finally, by choosing $\varepsilon$ smaller if
necessary, we can also ensure
$t_i^0+\varepsilon\delta_i\geq 0$ for every
$i\in N_k$. Since $\delta \in \Delta$, the aggregate
resource constraint is preserved for every
$\varepsilon$. Consequently, $z^\varepsilon$ is a feasible
allocation outside the core that cannot be
blocked by any unbalanced coalition. \(\heartsuit\)

Nevertheless $z^\varepsilon$ lies outside the core. To see this
directly, there must exist a cross-role pair $(1_r,2_s)$ with
$\delta_{1_r}+\delta_{2_s}<0$. Otherwise all $k^2$ pairwise
sums would be nonnegative; their sum is zero by the definition
of $V$, so all would be zero. This would force all role-1
coordinates to equal a common constant and all role-2
coordinates its negative; the two zero-sum constraints would
then imply $\delta=0$, a contradiction. For the pair with a
negative sum,
\[
 v(z_{1_r}^\varepsilon)+v(z_{2_s}^\varepsilon)
 =1+\varepsilon(\delta_{1_r}+\delta_{2_s})
 <1=v(1,1),
\]
so Lemma~\ref{lem:block} shows that this pair blocks
$z^\varepsilon$. Hence $z^\varepsilon$ is non-core but is
unblocked by every coalition in $\mathcal C$, contrary to
the fact that \(\mathcal C\) is an informational covering.

It follows that $\mathcal C$ must contain at least
$d+1=2k-1$ balanced coalitions. Along with Step 1, this yields
$|\mathcal C|\ge(2k-1)+2=2k+1$. The family $\mathcal F_k$ attains the bound, completing the proof.\medskip

\textbf{(iii)} The statement holds straightforwardly when \(k = 1\). For \(k \geq 2\), we prove that, for a suitable equal-treatment Pareto allocation
just \emph{below} the lower endpoint $c_k^-$, only coalitions of
profile $(k,k-1)$ can block. We then use the symmetric argument
just \emph{above} $c_k^+$.

\medskip\noindent
\emph{Step 1: The profile $(k,k-1)$ is indispensable.} Recall that, given some \(z(c)\), a coalition with profile \((\ell, m)\) can block it if and only if \(v(\ell, m) > \ell c + m (1-c)\). For any profile $\ell>m$, the is
equivalent to
\begin{equation}\label{eq:acondition}
  c<a_{\ell,m},\qquad
  a_{\ell,m}:=\frac{v(\ell,m)-m}{\ell-m}.
\end{equation}

We first establish that
\begin{equation}\label{eq:unique-lower}
  a_{\ell,m}<c_k^-
  \qquad\text{if }\ell>m\text{ and }(\ell,m)\ne(k,k-1).
\end{equation}
Recall that $\phi(t):=v(t,1)$ for $t\geq0$ and
\[
  q_+(t):=\frac{\phi(t)-\phi(1)}{t-1}
  =\frac{\phi(t)-1}{t-1},\qquad t>1.
\]
The function $\phi$ is strictly concave. 
Consequently, $\phi(t)$'s secant slope $q_+(t)$ is strictly decreasing in $t>1$. If $m>0$, homogeneity gives
\begin{equation}\label{eq:a-qplus}
 a_{\ell,m}=q_+\!\left(\frac{\ell}{m}\right).
\end{equation}
Since $\ell$ and $m$ are integers and
$1\leq m<\ell\leq k$,
\[
 \frac{\ell}{m}\geq \frac{k}{k-1},
\]
with equality only at $(\ell,m)=(k,k-1)$. Hence
\[
 a_{\ell,m}
 \leq q_+\!\left(\frac{k}{k-1}\right)
 =v(k,k-1)-(k-1)=c_k^-,
\]
and the inequality is strict except for $(k,k-1)$.
If $m=0$, homogeneity gives $a_{\ell,0}=v(1,0)$, while
strict superadditivity yields
\[
 v(k,k-1)>v(k-1,k-1)+v(1,0)
 =(k-1)+v(1,0).
\]
Thus $a_{\ell,0}<c_k^-$ as well, proving \eqref{eq:unique-lower}.

Because only finitely many profiles are possible, set
\[
 \alpha_k:=
 \max_{\substack{0\leq m<\ell\leq k\\(\ell,m)\ne(k,k-1)}}
 a_{\ell,m}<c_k^-.
\]
Choose
\begin{equation}\label{eq:c-low}
  \max\{0,\alpha_k\}<c<c_k^-.
\end{equation}
Then $z(c)$ is feasible and outside the core. Balanced coalitions cannot block $z(c)$,
because in that case $v(\ell, m) = \ell c + m (1-c) =\ell$. No coalition with $\ell<m$ can block: its nonblocking slack is affine in $c$ with negative slope and is nonnegative at
$c_k^-$ (a core allocation), hence is strictly positive for
$c<c_k^-$. Finally, by \eqref{eq:acondition} and \eqref{eq:c-low},
no coalition with $\ell>m$ and profile different from
$(k,k-1)$ can block. Coalitions of profile $(k,k-1)$ do block,
since their threshold is $c_k^-$. Therefore \emph{only} coalitions
with profile $(k,k-1)$ block $z(c)$. The sufficiency of
$\mathcal C$ forces it to contain at least one such coalition.

\medskip\noindent
\emph{Step 2: The profile $(k-1,k)$ is indispensable.}
For $\ell<m$, \(z(c)\) is blocked by a coalition with profile \((\ell, m)\) if and only if 
\begin{equation}\label{eq:bcondition}
  c>b_{\ell,m},\qquad
  b_{\ell,m}:=\frac{m-v(\ell,m)}{m-\ell}.
\end{equation}
Define
\[
 q_-(t):=\frac{\phi(1)-\phi(t)}{1-t}
 =\frac{1-\phi(t)}{1-t},\qquad 0\leq t<1.
\]
Strict concavity of $\phi$ implies that $q_-$ is strictly
decreasing in $t$. By homogeneity,
\begin{equation}\label{eq:b-qminus}
 b_{\ell,m}=q_-\!\left(\frac{\ell}{m}\right).
\end{equation}
For integers $0\leq\ell<m\leq k$, we have
\[
 \frac{\ell}{m}\leq\frac{k-1}{k},
\]
with equality only for $(\ell,m)=(k-1,k)$. It follows that
\begin{equation}\label{eq:unique-upper}
 b_{\ell,m}>c_k^+
 \qquad\text{if }\ell<m\text{ and }(\ell,m)\ne(k-1,k),
\end{equation}
since
\[
 q_-\!\left(\frac{k-1}{k}\right)
 =k-v(k-1,k)=c_k^+.
\]
Set
\[
 \beta_k:=
 \min_{\substack{0\leq\ell<m\leq k\\(\ell,m)\ne(k-1,k)}}
 b_{\ell,m}>c_k^+.
\]
Choose
\begin{equation}\label{eq:c-high}
 c_k^+<c<\min\{1,\beta_k\},
\end{equation}
$z(c)$ is feasible and outside the core. Balanced coalitions cannot block it.
No coalition with $\ell>m$ can block, since its nonblocking slack
is affine with positive slope and is nonnegative at $c_k^+$.
By \eqref{eq:bcondition} and \eqref{eq:c-high}, no coalition with
$\ell<m$ and profile different from $(k-1,k)$ can block.
Coalitions with profile $(k-1,k)$ do block, because their
critical value is $c_k^+$. Therefore only this profile can block
$z(c)$, so $\mathcal C$ must contain at least one coalition of
profile $(k-1,k)$.

Finally, a coalition with profile $(k,k-1)$ has the form
$N_k\setminus\{2_s\}$ for some $s$, and a coalition with profile
$(k-1,k)$ has the form $N_k\setminus\{1_r\}$ for some $r$.
This proves the claim.
\end{proof}

\end{document}